\documentclass[english,prb,amsmath,twocolumn,amssymb,nofootinbib,superscriptaddress,showpacs,floatfix,longbibliography]{revtex4-2}

\usepackage[english]{babel}

\usepackage[utf8]{inputenc}
\usepackage{latexsym}
\usepackage{braket}	
\usepackage{amsmath,amssymb,amsfonts,textcomp,latexsym,pb-diagram,amsopn,bm}
\usepackage{graphicx,xcolor}
\graphicspath{{pictures/}}
\DeclareGraphicsExtensions{.pdf,.png,.jpg,.eps}
\usepackage[breaklinks,colorlinks,bookmarks=false,citecolor=blue,linkcolor=red,urlcolor=blue]{hyperref}
\usepackage{url}
\usepackage{hyperref}
\usepackage{tabularx}
\newcolumntype{L}{>{\raggedright\arraybackslash}X}
\usepackage{epstopdf}

\begin{document}  

\title{ Universal separability criterion for arbitrary density matrices \\ 
from causal properties of separable and entangled quantum states }

\author{Gleb A. Skorobagatko*}

\affiliation{%
\mbox{%
Institute for Condensed Matter Physics of National Academy of Sciences of Ukraine,
  Svientsitskii Str.1,79011 Lviv, Ukraine%
  } \\
  *Correspondence to: gleb.a.skor@gmail.com}


\begin{abstract}
\begin{small}

General physical background of famous Peres-Horodecki positive partial transpose (PH- or PPT-) separability criterion is revealed. Especially, the physical sense of partial transpose operation is shown to be equivalent to what one could call as the \textit{"local causality reversal"} (LCR-) procedure for all separable quantum systems or to the \textit{uncertainty in a global time arrow direction} in all entangled cases. Using these universal causal considerations brand new general relations for the heuristic causal separability criterion have been proposed for arbitrary $  D^{N} \times D^{N}$ density matrices acting in $ \mathcal{H}_{D}^{\otimes N} $ Hilbert spaces which describe the ensembles of $ N $ quantum systems of $ D $ eigenstates each. Resulting general formulas have been then analysed for the widest special type of one-parametric density matrices of arbitrary dimensionality, which model a number of equivalent quantum subsystems being \textit{equally connected} (EC-) with each other to arbitrary degree by means of a single entanglement parameter $ p $. In particular, for the family of such EC-density matrices it has been found that there exists a number of $ N $- and $ D $-dependent \textit{separability (or entanglement) thresholds}  $ p_{th}(N,D) $ for the values of the corresponded entanglement parameter $ p $, which in the simplest case of a qubit-pair density matrix in $ \mathcal{H}_{2}  \otimes \mathcal{H}_{2} $ Hilbert space are shown to reduce to well-known results obtained earlier independently by Peres \cite{5} and Horodecki \cite{6}. As the result, a number of remarkable features of the entanglement thresholds for EC-density matrices has been described for the first time. All novel results being obtained for the family of arbitrary EC-density matrices are shown to be applicable for a wide range of both interacting and non-interacting (at the moment of measurement) multi-partite quantum systems, such as arrays of qubits, spin chains, ensembles of quantum oscillators, strongly correlated quantum many-body systems with the possibility of many-body localization, etc. 
\end{small}
\end{abstract}

\maketitle
\begin{normalsize}

\section{Introduction}

Entanglement of quantum states is the basic quantum phenomenon taking place in composite quantum systems of different nature, which refers to hidden interconnections between quantum dynamics and measurement outcomes for quantum states of system's constituent parts \cite{1,2,3,4,13,14,15,19,20,30,34}. Hence, quantum entanglement plays a crucial role in the majority of important applications of quantum mechanics including quantum computations protocols \cite{1,2,3,22,23} , quantum measurement and non-locality of quantum correlations between artificially prepared quantum states \cite{1,2,3,4,13,14,15,19,20}. Especially, preparation, driven evolution and driven interaction between different subsystems of any isolated quantum system as well as the subsequent either strong or weak quantum measurement of entangled quantum states of subsystems are all the essential parts of any quantum computation protocols based on the hardware represented by a composite quantum system given \cite{20,22,23}. Obviously, since any quantum system subjected to external manipulations is open, all the above mentioned manipulations can take place only at the level of the initially prepared density matrix $ \hat{\rho}_{i}(0) $ of a given composite quantum system, whereas different measurement outcomes within given quantum computation protocol allow for the reconstruction of system's density matrix $ \hat{\rho}(t) $ at different stages of its transformation including the final quantum state of the system described by density matrix $ \hat{\rho}_{f}(\tau) $ after time interval $ \tau $ of interest. 

Thus, in order to control the degree of entanglement and/or the effect of interactions between the subsystems of a given quantum system during its time evolution, it is necessary to find proper qualitative measures of the entanglement (or vice versa, separability), which would be encoded in the matrix elements of a given quantum density matrix describing the entire quantum system at arbitrarily chosen moment of time \cite{1,3,5,6,7,8}. As the result, a number of different entanglement (or  separability) criteria \cite{3,5,6,7,8,28,29,31,32,34,35} has been developed for finite-dimensional quantum density matrices describing composite open quantum systems which are defined in finite-dimensional Hilbert spaces  (for the review on different entanglement measures one can see e.g. well-known review article of Ref.\cite{3}). 

Technically, most of the existing separability criteria for arbitrary density matrices are based on the positivity of different (mostly quadratic but sometimes more complicated) forms   combined from the density matrix elements involving so-called an "entanglement witness" operator (see e.g. Refs.\cite{3,5,6,9,10,28,29,31,32}) - , which one may think of as the eigenvalues of certain new matrices obtained from a density matrix given after definite discrete transformations performed on the latter \cite{3,5,6,9,10,28,29,31,32}. This includes also probably the most famous one: Peres-Horodecki \textit{positive partial transpose-}(or simply PPT-) separability criterion for density matrices acting in $ 2 \otimes 2 $ and $ 2 \otimes 3 $ Hilbert spaces\cite{3,5,6}. Here one can consider the operator of density matrix partial transposition (or equally the operator of virtual quantum transition-virtual spin-flip, see below) as such a sort of the "entanglement witness" \cite{31}. 

However, except very special Heisenberg uncertainty-based approaches \cite{7,29,31,32}, where one analyses correlations in the Heisenberg uncertainties in the measurement outcomes for different entanglement witnesses and the approach to separability in  terms of system bi-partitions  \cite{8,28,31,35} based on the idea of EPR correlations and Bell inequalities violation, there has been no any general physical idea behind such the entanglement witnesses construction yet. This implies the main problem for all existing approaches to the separability criteria: \textit{the entanglement witness operator construction for any quantum system is not unique}, that is why the estimations of the entanglement degree in given quantum system or ensemble one makes by using different separability criteria can differ from each other substantially \cite{3,5,6,28,29,31,32,35}.  Exactly because of this a lack of understanding of a physical meaning for underlying discrete transformations of density matrices behind the separability criteria  prevents any sufficient further development in this important direction of quantum physics. For example, no general  separability criteria is yet known for a widest possible class of density matrices describing either separable or entangled arbitrary $ N $-partite quantum systems of $ D $ dimensionality each, where arbitrary $ N $ is a number of identical subsystems in the given quantum system (ensemble) and arbitrary $ D $ is the dimensionality of the Hilbert space for each quantum subsystem of the ensemble.  

Hence, below I propose the original way to determine a new universal separability criteria for such a wide class of density matrices based on the analysis of underlying causal physical nature of a well-known Peres-Horodecki's (PH-) ppt-separability criterion \cite{5,6} which is also analysed in below. As the result, a novel physical background of causal nature for PH-criterion is uncovered and then the entirely new type of quite general \textit{causal} separability criteria is proposed on this theoretical platform. The latter is shown to be valid for arbitrary density matrices acting in $ \mathcal{H}_{D}^{\otimes N} $ Hilbert spaces with arbitrary (either finite or even infinite) integer numbers $ D $ and $ N $. (In order to shorten notations in below I will use the symbol $ D^{\otimes N}  $ instead of $ \mathcal{H}_{D}^{\otimes N} $ for the entire Hilbert space in which density matrices of interest are defined.) 

The central idea of a novel approach being proposed in below is to circumvent the problem of the ambiguity in the entanglement witnesses construction by analysing instead another general properties which distinguish between separable and entangled quantum states of arbitrary multi-partite quantum system in the most general case. The latter properties are encoded in different causal relations for the ignorance- (meaning the ignorance about the true quantum state of the system) and in virtual quantum transition probabilities for all separable and all entangled cases, correspondingly. These causal relations are shown to be intimately connected with general symmetry of arbitrary separable and arbitrary entangled quantum states with respect to the operation of \textit{local causality reversal} (for all possible separable states) or with respect to the uncertainty in a global time arrow direction (for all possible entangled states).  

The  novel universal causal approach developed is shown to have common consequences and applications of general formulas for new  causal separability criteria obtained are discussed including the prominent correspondence with a number of facts about the entanglement in complex quantum systems of different nature which have been already known \cite{5,6,12,14,16,17,19,21}. Possible applications of general results obtained are seemed to be very broad: almost every experimentally accessible composite quantum system which can be prepared, manipulated and measured (either strongly or weakly) \cite{2,4,11,12,13,14,15,20,21,22,23} as well as any strongly correlated quantum system of many interacting identical particles e.g. \cite{16,17,19,20,21,27} provides us with its time-dependent density matrix elements which encode either separable or entangled quantum states of its constituent quantum subsystems. Therefore, the theory developed in below is able to provide proper signal processing in many schemes of quantum detection (see e.g.\cite{24,25}) including e.g. non-invasive quantum detection of quantum vibrations in photonic and optomechanical devices \cite{4,13} and the networks of the solid state- and superconducting qubits \cite{20,23}. This way new universal causal separability criterion is able to determine whether the signal output from any complicated quantum hardware encodes separable or entangled quantum state of underlying quantum computations network \cite{22,30,34,35}.      

To proceed in this direction, first, one can notice that all matrix elements of arbitrary density matrix of interest should be thought of as ones taken at certain moment of time, i.e. these matrix elements by themselves are results of quantum evolution of a certain initially prepared density matrix if to understand such evolution as one including all the manipulations, projective measurements, preparations, etc. performed on a given density matrix in the past. Obviously, the same should also be valid for any initial density matrix being discrete transformed e.g. for ones being \textit{partially transposed} (or equally, pt-transformed). Therefore, the guiding idea here will be to connect well-known \cite{3,5,6,9,10} discrete partial transpose (pt-) transformations of the  most simple density matrices of composite bi-partite quantum systems in $ 2 \otimes 2 $ Hilbert space with certain general features in their initial preparation in the past depending on the fact whether this initially prepared quantum state was product-state  $ \vert A \rangle  \otimes \vert B \rangle$  of two constituent parties ($ A $ and $ B $) of the system or not at the moment of system  "preparation". Then it becomes possible to encode such two types of preparation scenarios in properties of definite matrix elements combination for arbitrary $ 4 \times 4 $-dimensional density matrix given \cite{5,6}. This observation, in turn, will make one possible to apply the same logic in order to derive heuristic formulas for the separability (or equally, entanglement) criteria of the same physical meaning being valid for the widest class of density matrices of arbitrary dimensionality defined in $ D^{\otimes N} $ Hilbert  space. This represents a guideline taken in this research.

The paper is organized as follows. In Section II the causal nature of Peres-Horodecki ppt-separability criterion is revealed. In Section III the universal causal separability criterion for density matrices is formulated in its most general form. In Section IV the configuration description of a generic density matrix in $ D^{\otimes N} $ Hilbert space is given. In Section V the novel universal causal separability criterion is formulated in terms of configurations (of subsystems' orthogonal base quantum states). In Section VI the general form of the universal causal separability criterion being found  is tested on the particular wide category of one-parametric \textit{equally connected} (or EC-) density matrices from the $ D^{\otimes N} $ Hilbert space with arbitrary $ D $ and $ N $, as well, in this section the discussion on the physical consequences of all obtained results for a wide range of realistic multi-partite quantum systems and ensembles is performed. In Section VIII the conclusions are given.    

\bigskip

\section{Causal nature of ppt- separability criterion for two-qubit density matrix} 

To begin with let us define a four-dimensional density matrix $ \rho(t) $ for the simplest bipartite quantum system (say, a pair of qubits e.g. a pair of $ 1/2 $ spins ) defined in the $ 2 \otimes 2 $ Hilbert space  as $ \rho_{\mu m \vert \nu n} $ at certain moment $ t $ , where all greek indices denote quantum state of one subsystem (say, A-subsystem) of given quantum system, whereas all latin indices stand for its another subsystem(say, B subsystem) \cite{5,6}. Here each (either greek or latin) index runs only $ 2 $ possible values, say, $ 1 $ and $ 2 $ \cite{5,6}. Since any density matrix represents certain transformation  of base ket-vectors which describe given quantum system in the Hilbert space, it is convenient to introduce a vertical slash in lower indices which separates the indices of initial base ket-vectors  (ones on the right hand-side (r.h.s.) from the slash) which given density matrix transforms to certain "new" ket-vectors which are denoted by corresponding indices on the left hand-side (l.h.s.) from the slash. In such notations a partial transposition  (i.e. pt-transformation) of given density matrix takes place only in the Hilbert subspace of one among two subsystems reads $ \rho^{T_{1}}_{\mu m \vert \nu n}= \rho_{\mu n \vert \nu m}$ (here superscript $ T_{1} $ stands for the partial transposition of a given matrix with respect to Hilbert subspace of one party \cite{5,6}). Thus, Peres-Horodecki (or ppt-) separability criterion states that \textit{iff all eigenvalues of the pt-transformed density matrix are non-negative then given density matrix is separable, otherwise, it is entangled} \cite{5,6}. This actually means that in the first (separable) case a quantum state of a given bipartite quantum system was a product state of its two subsystems at the moment of system preparation, while in the second (entangled) case one could, in principle, "distil"  certain entangled quantum state \cite{3,4,13,14,15} by some manipulations performed on the entire quantum system of both constituent parties given bipartite quantum system has started to evolve from (for some examples of the entanglement distillation procedure one can see e.g. Refs.\cite{1,2,3,4,20,21} ). 

\subsection{ Mapping of the ppt-criterion  on the  \textit{Local Causality Reversal (LCR-)} procedure for density matrices in $ 2 \otimes 2 $ Hilbert space.}

From the above description of the ppt-criterion it is clear that the latter has to do with the "history" of preparation and further evolution of a composite quantum system given density matrix describes \cite{3,7,8,9,10}. Remarkably, from the definition of ppt-criterion it becomes clear that  \textit{pt-transformation of a given density matrix represents a procedure which preserves a probabilistic meaning of density matrix in the separable case and destroys such a meaning in the entangled case}. The latter consequence becomes a bit more pronounced e.g. in the development of the continuous variable modification \cite{7,8,34,35} of ppt-separability criterion, where one deals with correlations of the quantum Heisenberg uncertainties in the measurement outcomes for gaussian continuous variables which describe quantum states shared between two parties in bipartite quantum systems \cite{5,6,7,8}. Thus, one may ask what kind of physical background can hide behind such a partial transpose procedure in the simplest case of bipartite density matrices in $ 2 \otimes 2 $ Hilbert space ? - Actually, it is not too hard to answer. Indeed, the partial transpose (pt-) changes only those off-diagonal matrix elements from the off-diagonal blocks which act only in one among two 2-dimensional subspaces of the entire 4-dimensional Hilbert space given density matrix is defined \cite{5,6}. This means the pt-transformation of density matrix interchanges probability amplitudes for one subsystem to transfer from its $ n $-th to $ m $-th  base quantum state in certain orthonormal basis chosen, while all probability amplitudes of quantum transitions of any kind  in another subsystem (to transfer from its $ \nu $-th to $ \mu $-th quantum state and vice versa) remain intact. One may convince in the latter fact just by tracing out the entire density matrix partially, in the Hilbert subspace(s), where pt-transformation of its matrix elements takes place.  In other words, under pt-transformation all quantum transitions in one subsystem from its $ n $-th to its $ m $-th quantum state should be replaced by the \textit{inverse} transitions from its $ m $-th to its $ n $-th quantum state (and vice versa).  But, evidently, \textit{the same result can be achieved in the hypothetical situation, where the arrow of time would be reversed only locally in one subsystem of a given bipartite quantum system, while in another subsystem no such time reversal occurs}. Interestingly, initial formulations of this important observation about the nature of pt-transformation were made by several researchers earlier \cite{9,10}. However, previous studies \cite{9,10} on this important aspect of ppt-separability criterion concerned only the simplest special cases of  density matrices acting in $ 2 \otimes 2 $ and $ 2 \otimes 3 $ Hilbert spaces \cite{9} including the $ 2 \otimes 2 $ case of continuous variables \cite{10} and both studies did not outline any general approach from this peculiarity. In what follows the most general causal approach to separability/entanglement criterion will be demonstrated for the first time and it will be extensively studied for the most general type of density matrices acting in Hilbert space of arbitrary dimensionality.  

Therefore, without the loss of generality, one can claim that Peres-Horodecki pt-transformation\cite{5,6} of quantum ensemble's density matrix has physical meaning of the \textit{Local Causality Reversal} (or simply LCR-)operation being applied to the local time arrow in one among the constituent subsystems of a given composite quantum system. Notice, since the term "causality" can make sense locally (as any connection between the events taking place in any system), it fits our purposes better than the term "time arrow" (since the latter is global in its usual meaning). The same argument can be applied to the density matrices of arbitrary dimensionality. Therefore, standard ppt-separability criterion \cite{5,6} claims that all separable density matrices being pt-transformed have non-negative eigenvalues. This means that respective pt-transformed density matrix $ \hat{\rho}^{T_{1}}(t) $ maintains its probabilistic meaning and therefore the LCR-transformed density matrix $ \hat{\rho}^{\textit{LCR}_{1}}(t) $ also makes sense in the separable case. On the other hand, since the standard ppt-separability criterion \cite{5,6} states that all those density matrices describing entangled quantum states of the system after the pt-transformation will have negative eigenvalues, one can conclude that the \textit{LCR operation does not make sense for one among two quantum subsystems if they are entangled even to the smallest degree because in this case the corresponding LCR-transformed density matrix of the entire system loses its probabilistic sense}. Hence, in all entangled cases respective LCR -transformed density matrix $ \hat{\rho}^{\textit{LCR}_{1}}(t) $ simply does not exist. This means that for arbitrary density matrices $ \hat{\rho}(t) $ describing arbitrary multi-partite quantum system one can claim that  

\begin{align}
\begin{split}  
\ \ \hat{\rho}^{\textit{LCR}_{m}}(t) =\left\{ 
 \begin{matrix}
 \hat{\rho}^{T_{m}}(t),  \ \ -separable \ case \ \\  
\\ 
  \varnothing,  \ \  -entangled \ case \
        \end{matrix} \right.
 \end{split}
\end{align}

where the superscript $ \textit{LCR}_{m}  $ means local causality reversal operation performed on $ m $ among $ N $ quantum subsystems given quantum system consists of, while the superscript $ T_{m} $ denotes usual partial transpose (pt-
) operation performed on matrix elements of a given density matrix in $ m $ among its $ N $ Hilbert subspaces in the case of $ N $-partite quantum system given density matrix encodes. Notice, in what follows we will use mostly $ T_{m} $ notation for any $ \textit{LCR}_{m}  $  operations in the separable case  in order just t.o shorten matrix superscripts.

 Obviously, Eq.(1) opens the road to generalisation of ppt-separability/entanglement criterion on arbitrary quantum composite systems and ensembles. On one hand, within the terminology proposed in the above a true physical background of Peres-Horodecki separability criterion is now revealed: \textit{Only when quantum states of constituent subsystems are separable from  each other the LCR-procedure in any constituent subsystem is possible independently from the other constituent subsystems/parties of the entire quantum system.} Nothing prevents this situation from being experimentally modelled and observed, hence, corresponded pt- (or equally, LCR-) transformed density matrix should preserve its probabilistic (or equally density matrix-) meaning due to the positivity of its eigenvalues ( ppt-separability criterion). On the other hand, for any degree of entanglement between quantum states of constituent subsystems \textit{the possibility of the LCR- procedure on any fraction of subsystems being entangled with the rest of subsystems in a given quantum system would automatically presume the uncertainty of a global time arrow for the entire (entangled) quantum system}. \textit{The latter uncertainty, in turn, forbids any well-defined independent temporal evolution of any among system's constituent parties being entangled with each other, because the opposite situation (i.e. the possibility of such independent time evolution of any entangled subsystems) would automatically imply  e.g. the possibility of a superluminal signalling between the LCR-transformed subsystems and the rest of the ensemble as a result of the entanglement between them, which, of course, is not the case in nature. Thus, the LCR (or pt-) procedure performed on density matrices which encode entangled quantum states can never lead to any observable consequences resulting in the presence of negative numbers among the eigenvalues of such LCR transformed density matrices. } Interestingly, the validity of the above statement has been recently supported indirectly by the general claim being proven in Ref.\cite{33}  about the impossibility of any LOCC- (local operation assisted by classical communication) transformations performed on any constituent parties of any multi-partite entangled quantum system. 

\subsection{Reformulation of the ppt-separability criterion in terms of determinants for pt-transformed  $ 2 \otimes 2 $  density matrices.}

In the simplest $ 2 \otimes 2 $ case it is easy to show \cite{5,6} that ppt-separability criterion is equivalent to the requirement of the positivity for both following determinants 

\begin{eqnarray} 
W_{1}= \rho^{T_{1}}_{11 \vert 11} \rho^{T_{1}}_{22 \vert 22} - \rho^{T_{1}}_{11 \vert 22} \rho^{T_{1}}_{ 22 \vert 11}  \nonumber \\
W_{2}= \rho^{T_{1}}_{12 \vert 12} \rho^{T_{1}}_{21 \vert 21} - \rho^{T_{1}}_{12 \vert 21} \rho^{T_{1}}_{ 21 \vert 12}
\end{eqnarray}

constructed from the matrix elements of a given pt-transformed $ 2 \otimes 2 $ density matrix $ \hat{\rho}^{\textit{T}_{1}} $ of arbitrary bipartite quantum system, where each among its two subsystems "lives" in the 2-dimensional Hilbert subspace ( here $ \mu,\nu,n,m = 1,2 $ ). If both $ W_{1,2} $ are non-negative then corresponded density matrix describes separable bipartite quantum system, whereas if either $ W_{1} $ or $ W_{2} $ is negative then corresponding bipartite quantum system is entangled \cite{5,6} . Evidently, Eqs.(2) "translate"  ppt-separability/entanglement criterion on the language of certain  probabilistic measures which characterize the result of preparation, evolution and interaction between the parties of given bipartite quantum system. Thus, in what follows we will take Eqs.(2) as the alternative definition of ppt-separability/entanglement criterion and in such a way will uncover the physical background of Eqs.(2) in the "history" of the pt-transformed quantum system. This line of  considerations, in turn, will enable us to formulate a much more general criterion of separability (or entanglement) for generic multi-partite $ D^{\otimes N} $ quantum system.  

First of all, let us generalize Eqs.(2) by using new shorthand notations for the indexing. Namely, let in our bipartite case the symbol $ \lbrace k \rbrace_{j}= \mu_{j} m_{j} $  where $ j=1,2 $, denotes a particular ($ j $-th) composition of two base states which characterizes a particular quantum state of the entire bipartite quantum system one is able to know with certainty by measuring strongly the eigenstates of certain observable $ k $ which has $ D=2 $ eigenvalues in each among $ N=2 $ constituent subsystems of a given quantum system. (Let the first index in such composition denotes base state of the subsystem A and the second index - stands for the state of subsystem B.) Further, let the symbol $ \lbrace \bar{k} \rbrace_{j}= \nu_{j} n_{j} $ (where $ \nu \neq \mu $ and $ n \neq m $) defines such the composition of subsystems' base quantum states where each element from this composition  is orthogonal to the corresponded element from the chosen composition $ \lbrace k \rbrace_{j} $. In what follows we call $ \lbrace \bar{k} \rbrace_{j} $ as the \textit{completely orthogonal composition of base states} for any particular composition of base states $ \lbrace k \rbrace_{j} $. For example, if, independently for each party of the system, one has $ \langle \mu_{j} \vert \nu_{j} \rangle=0 $ and $ \langle m_{j} \vert n_{j} \rangle=0 $  to be fulfilled then we denote  $ \lbrace k \rbrace_{j}= \mu_{j} m_{j} $ and $ \lbrace \bar{k} \rbrace_{j}= \nu_{j} n_{j} $ , meaning that $ \vert \lbrace k \rbrace_{j} \rangle= \vert \mu_{j} \rangle \vert m_{j} \rangle $ and $ \vert \lbrace \bar{k} \rbrace_{j} \rangle = \vert \nu_{j} \rangle \vert n_{j} \rangle $  with the evident consequence $ \langle \lbrace k \rbrace_{j} \vert  \lbrace \bar{k} \rbrace_{j} \rangle=\langle \lbrace \bar{k} \rbrace_{j} \vert \lbrace k \rbrace_{j} \rangle=0 $ for any $ j=1,2 $ with $ D=2 $ being the number of base states for each subsystem (since in general case we will presume our system to be consisted of $ N $ identical quantum subsystems with $ D $  base states each). 
After that one can generalize both equations in (2) for  bipartite quantum system with $ D=N=2 $ as follows

\begin{eqnarray} 
W^{j}= \rho_{\lbrace k \rbrace_{j} \vert \lbrace k \rbrace_{j}} \rho_{\lbrace \bar{k} \rbrace_{j} \vert \lbrace \bar{k} \rbrace_{j}} - \rho^{T_{1}}_{\lbrace k \rbrace_{j} \vert \lbrace \bar{k} \rbrace_{j}} \rho^{T_{1}}_{ \lbrace \bar{k} \rbrace_{j} \vert \lbrace k \rbrace_{j}},
\end{eqnarray}

for $ j=1,2 $. Evidently, Eq.(3) automatically reproduces two Eqs.(2). Hence, according to ppt separability criterion \cite{5,6},  given bipartite quantum system is separable for $ W^{j} \geq 0 $ and entangled in the case where $ W^{j} < 0 $ for any $ j=1,2 $. The remarkable advantage of the form (3) of Eqs.(2) is that the notations chosen for Eq.(3) allow one to construct similar mathematical forms for any number of dimensions in the Hilbert space of a generic composite quantum system. As we will see in what follows form (3) can be straightforwardly modified on the multi-partite case of $ \mathcal{H}_{D}^{\otimes N} $ with $ N $ and $ D $ being arbitrary natural numbers.

One can see that, remarkably, Eq.(3) incorporates only two types of probabilistic quantities with clear physical sense each: 1)$ \rho^{T_{1}}_{\lbrace k \rbrace_{j} \vert \lbrace k \rbrace_{j}}=\rho_{\lbrace k \rbrace_{j} \vert \lbrace k \rbrace_{j}} $ and $  \rho^{T_{1}}_{\lbrace \bar{k} \rbrace_{j} \vert \lbrace \bar{k} \rbrace_{j}} = \rho_{\lbrace \bar{k} \rbrace_{j} \vert \lbrace \bar{k} \rbrace_{j}} $  which has a sense of the \textit{probability} for the entire  \textit{either LCR-transformed or not} quantum system to be in the $ \lbrace k \rbrace_{j} $ -th composition of base quantum states of all its subsystems (and the probability to be in the $ \lbrace \bar{k} \rbrace_{j}$-th  composition of all those subsystems' base states being all orthogonal to all states from the  $ \lbrace k \rbrace_{j} $-th composition, correspondingly); and  2) $ \rho^{T_{1}}_{\lbrace k \rbrace_{j} \vert \lbrace \bar{k} \rbrace_{j}} $ (and $ \rho^{T_{1}}_{ \lbrace \bar{k} \rbrace_{j} \vert \lbrace k \rbrace_{j}} $ ) - according to the standard quantum mechanical definition - represents just a \textit{quantum transition probability amplitude} for the entire \textit{pt-transformed} quantum system to transfer from the particular ($ j $-th) composition $ \lbrace \bar{k} \rbrace_{j} $ of its base states to another $ \lbrace k \rbrace_{j} $  composition of its base states (and vice versa)  being  completely orthogonal to the former. One can interpret such a common kind of quantum transitions between different base quantum states of  the system  as the result of a certain external perturbation and/or state preparation manipulations being performed on the entire quantum system.  

\section{Universal causal separability criterion for density matrices}

As it has been already mentioned in the Introduction there are lots of available separability criteria \cite{3,28,29,31,32} based on different entanglement witnesses which exploit the quadratic forms being similar to ones of the type of Eqs.(2,3) but at this point an interesting question can be formulated: what can be the most general physical meaning of Eqs.(2,3) in the context of preparation of either separable or entangled quantum state of generic bipartite quantum system?  In order to answer it all the constituent terms of our Eq.(3) will be analysed with respect to their probabilistic meaning in what follows .

\subsection{Ignorance- and virtual quantum transition probabilities}

 First of all, it is quite obvious that the probability $ \rho_{\lbrace k \rbrace_{j} \vert \lbrace k \rbrace_{j}} \rho_{\lbrace \bar{k} \rbrace_{j} \vert \lbrace \bar{k} \rbrace_{j}} $ can be treated as the \textit{measure of observer's ignorance} about the quantum state of generic quantum system in $ 2 \otimes 2 $ Hilbert space. In other words, the latter measure represents estimation of the probability for the situation in system's history where the observer cannot say which composition of system's base states will be realized in the upcoming moment of time irrespectively to the fact whether given system have been LCR-transformed or not before. Naturally, the latter can happen with any quantum system only at the moment of the "preparation" of its entire density matrix or during the period of  certain "contact interaction" between its constituent subsystems and evidently one can detect such stage either observing the system in the past or by "modelling" such the stage in system's history by means of a local causality reversal (LCR-) transformation (1) of the system.  Thus, we can denote this "observer's ignorance" probability in the case of  bipartite quantum system with $ D=N=2 $ as 

\begin{eqnarray} 
P^{(j)}_{\varnothing}=\rho_{\lbrace k \rbrace_{j} \vert \lbrace k \rbrace_{j}} \rho_{\lbrace \bar{k} \rbrace_{j} \vert \lbrace \bar{k} \rbrace_{j}}, 
\end{eqnarray}
 
where lower symbol $ \varnothing $ marks the fact that corresponding probability estimates the stage of system's preparation/projective measurement where the  $ j $-th combination  of eigenstates of certain observable $ k $ cannot be assigned to a given quantum system at the moment of its density matrix preparation (or during the contact interaction between its constituent subsystems). Naturally, Eq.(4) is invariant with respect to LCR operation because for the "ignorance" probability it does not mean due to which specific sort of subsystems' temporal evolution such the "ignorance" in system's quantum state has been occurred. 

The meaning of the next term $ \rho^{T_{1}}_{\lbrace k \rbrace_{j} \vert \lbrace \bar{k} \rbrace_{j}} \rho^{T_{1}}_{ \lbrace \bar{k} \rbrace_{j} \vert \lbrace k \rbrace_{j}} $ in Eq.(3) seems to be a bit more familiar: it is just the probability of a \textit{ virtual quantum transition} between certain $ \lbrace k \rbrace_{j} $ combination of base quantum states and another base states combination $ \lbrace \bar{k} \rbrace_{j} $  being completely orthogonal to the former. Such a sort of virtual quantum transition may occur either in the past of given quantum system at the stage of its preparation or, alternatively, during the process of the causality reversal when system undergo again the moment of its preparation from its past. Obviously, at any moment when such a virtual transition happens (for example, in the case of two spins one can treat the latter as virtual spin-flip), especially at the moment of given density matrix preparation, one cannot identify the entire quantum state of the system with respect to eigenvalues of observable $ k $ from the spectrum of eigenvalues of this observable being strongly measured. That is why during such a virtual quantum transition the observer \textit{remains ignorant} about the existence of $ j $-th combination of subsystems' base states with respect to eigenstates of measured observable $ k $ in given quantum system. Therefore, one can introduce corresponding virtual quantum transition probability in the bipartite quantum system with $ D=N=2 $ as follows
  
\begin{eqnarray} 
P^{(j)}_{\circlearrowright} = \rho^{T_{1}}_{\lbrace k \rbrace_{j} \vert \lbrace \bar{k} \rbrace_{j}} \rho^{T_{1}}_{ \lbrace \bar{k} \rbrace_{j} \vert \lbrace k \rbrace_{j}},   
\end{eqnarray}  
  
 where the lower symbol $ \circlearrowright $ of a clockwise-closed arrow symbolizes virtual transition between two orthogonal compositions of base states of the system which accompanies density matrix preparation- or measurement procedure and happens accordingly to chosen global time-arrow for the system (a clockwise direction).  

The above considerations together with Eqs.(4,5) give us some hints on the probabilistic sense of $ W^{j} $ quantities in Eq.(3) despite the latter are seemed to be less clear. To reveal it we should consider separately cases $ W^{j} \geq 0 $ (separable state of two subsystems) and $ W^{j} < 0 $ (entangled state of two subsystems). Let us introduce non-negative quantity $ \vert W^{j} \vert \leq 1 $ which, obviously, can have certain probabilistic sense. Then one can rewrite Eq.(3) in the following form

\begin{eqnarray} 
P^{(j)}_{\varnothing} = P^{(j)}_{\circlearrowright} + \vert W^{j} \vert   
\end{eqnarray}

for any separable and 

\begin{eqnarray} 
P^{(j)}_{\circlearrowright} = P^{(j)}_{\varnothing} + \vert W^{j} \vert    
\end{eqnarray}

for any entangled density matrix of any $ 2 \otimes 2 $ quantum system.

\subsection{Basic causal relation between the ignorance- and virtual quantum transition probabilities in the separable and entangled cases.}

Now in the framework of the standard probability theory it becomes clear that probability sums of Eqs.(6,7) represent certain "channel decompositions" for the situation of quantum state uncertainty ( $ \varnothing $ ) in the separable case - and for the situation of a virtual quantum transition ( $ \circlearrowright $ ) between two orthogonal compositions of system's base states in the entangled case to occur somewhere in system's history. On one hand, in the above it was shown that in the separable case the LCR procedure makes sense. On the other hand, in the separable case the determinant $ W^{j} = \vert W^{j} \vert $ is connected with eigenvalues of LCR- (or pt-) transformed initial density matrix. Taking this into account, one can conclude that Eq.(6) describes the fact that \textit{observer's impression about the uncertainty of system's quantum state} in the separable case during the system preparation/measurement procedure can be achieved in two independent ways: i) during the process of virtual quantum transition between two orthogonal combinations of subsystems' base states with respect to eigenvalues of observable $ k $ being measured for each subsystem or ii) during the same kind of virtual quantum transition though occurred in the process of local causality reversal (i.e. during the LCR- transformation ) in a given quantum system. Therefore, one can mark the probability of such a virtual quantum transition for $ j $-combination of subsystems' base quantum states in the LCR-transformed quantum system as $ P^{(j)}_{\circlearrowleft} = \vert W^{j} \vert = W^{j} $ . Here the lower symbol $ \circlearrowleft $ denotes the fact that a virtual quantum transition in the LCR-transformed separable quantum system (e.g. virtual spin-flip, or more generally quantum fluctuation of $ k $ observable) should realize the \textit{inverse causality} in the LCR- transformed subsystem(s) of a given quantum system (this fact is reflected in the anti-clockwise direction of a closed arrow in the corresponded subscript). Then Eq.(6) for the separable case can be represented in the following remarkable form

\begin{eqnarray} 
P^{(j)}_{\varnothing} = P^{(j)}_{\circlearrowright} + P^{(j)}_{\circlearrowleft}.   
\end{eqnarray}

In what follows we will see that Eq.(8) is \textit{universal} for any density matrices which describe separable quantum states since it encodes basic causal relations for preparation-related probabilities in all separable cases of composite quantum system evolution (see also Fig.1 for illustration).

In the entangled case $ W^{j} < 0 $, hence, as it has been already mentioned in the above the LCR procedure on such the entangled quantum system is impossible. Therefore, physical sense of corresponded decomposition of Eq.(7) should differ from one in the separable case. Though the sense of Eq.(7) should be still a certain decomposition of virtual quantum transition (say, virtual spin-flip or quantum fluctuation of observable measured) on two possible global causal orderings such the transition could obey in the entangled quantum system. As it follows from Eq.(7) one variant here is just the observer's ignorance about the true quantum state of the system at the moment of density matrix preparation. This possibility is represented by the first term in the sum from the r.h.s. of Eq.(7).  Whereas, another option (i.e. the second term in the sum of Eq.(7)) in the entangled case cannot be associated with any LCR procedure since the latter is impossible for the quantum system in the entangled state. However, such the impossibility has certain positive sense. There exists an intuitively clear consideration: the direction of a local time arrow of any subsystem in a bigger quantum system can be determined only relatively to a freely chosen direction of a local time arrow in another independent (or separable) subsystem of the entire quantum system. But on the other hand, in the case where all subsystems of a given quantum system are entangled with each other (even to the smallest degree) one is unable to choose any fixed direction of a local time arrows, whereas in such the case there exists only one "global" time arrow for the time evolution of the entire entangled quantum system. But the point is that the direction of such a global time arrow is impossible to detect within any virtual quantum transition in the entire entangled system. Therefore, Eq.(7) should mean that a virtual quantum transition in arbitrary entangled quantum system \textit{can be noticed} by the observer in two different ways: i) via observer's  "ignorance"  about the concrete ($ j $-th) composition of subsystems' base states while the system undergoes a virtual quantum transition (e.g. during the entangled density matrix preparation), which is described by the probability $ P^{(j)}_{\varnothing} $ and is associated with, say, "forward" direction of a global time arrow and ii) via the same kind of observer's ignorance which though takes place when a global time arrow changes its direction to the opposite (this is the analog of the LCR procedure for the entangled quantum system). Obviously, corresponded probability of observer's ignorance about the $ j $ composition of subsystems' base states in the entangled quantum system while the direction of a global time arrow is reversed can be denoted as  $ P^{(j)}_{- \varnothing} = \vert W^{j} \vert = - W^{j}$, since in the entangled case one has $ W^{j} < 0 $. Here the lower symbol $ - \varnothing $ marks the uncertainty in quantum state of a given entangled quantum system when the global time arrow is reversed.  All the above allows one to summarize Eq.(7) for the entangled case in the form (see also Fig.2 for the illustration)

\begin{eqnarray} 
P^{(j)}_{\circlearrowright} = P^{(j)}_{\varnothing} + P^{(j)}_{- \varnothing}.   
\end{eqnarray}

\begin{figure}
\includegraphics[height=13 cm,width=9 cm]{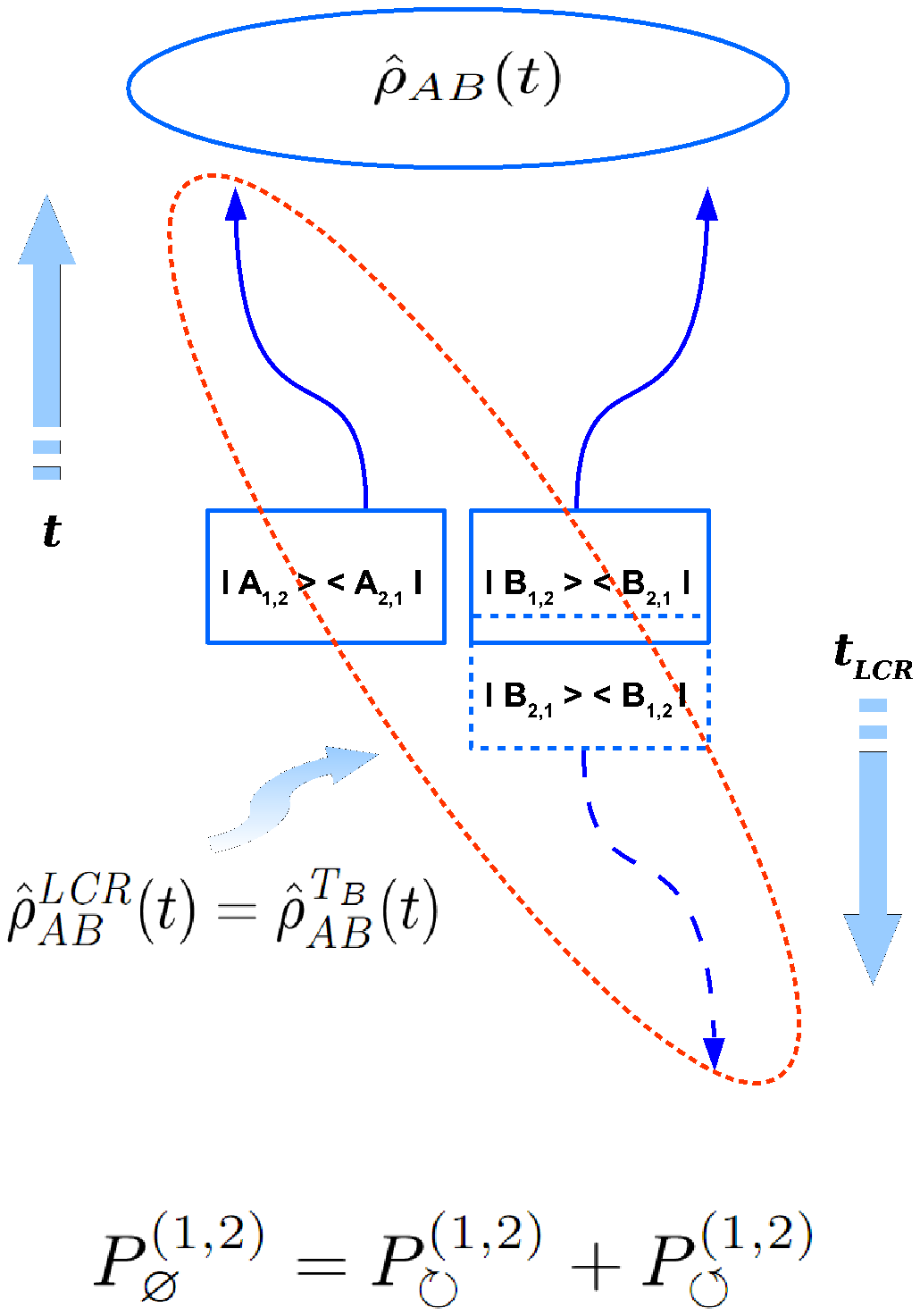}
\caption{Schematic picture of $ 4 \times 4 $ (two qubit- ) density matrix evolution in the \textit{separable case} with respect to the \textit{local causality reversal} (LCR-) procedure applied to $ B $ -subsystem which is possible in the separable case. Square blocks are supposed to mark a moment of a given density matrix preparation at $ t=t_{LCR}=0 $. Such preparation involves statistically independent virtual quantum transitions in both subsystems depicted by means of projectors in the blocks (here the solid blocks start forward time evolution while dashed block marks the start of locally causal-reversed evolution of B-subsystem) . Since all blocks  for  A- and B subsystems are disconnected from each other in the separable case, the LCR-procedure on each subsystem makes sense and hence any LCR- (or pt-) transformed separable density matrix $ \hat{\rho}^{\textit{LCR}}_{AB}(t) $ from $ 2 \otimes 2  $ Hilbert space maintains its probabilistic sense in such the case. }
\end{figure} 

\begin{figure}
\includegraphics[height=13 cm,width=9 cm]{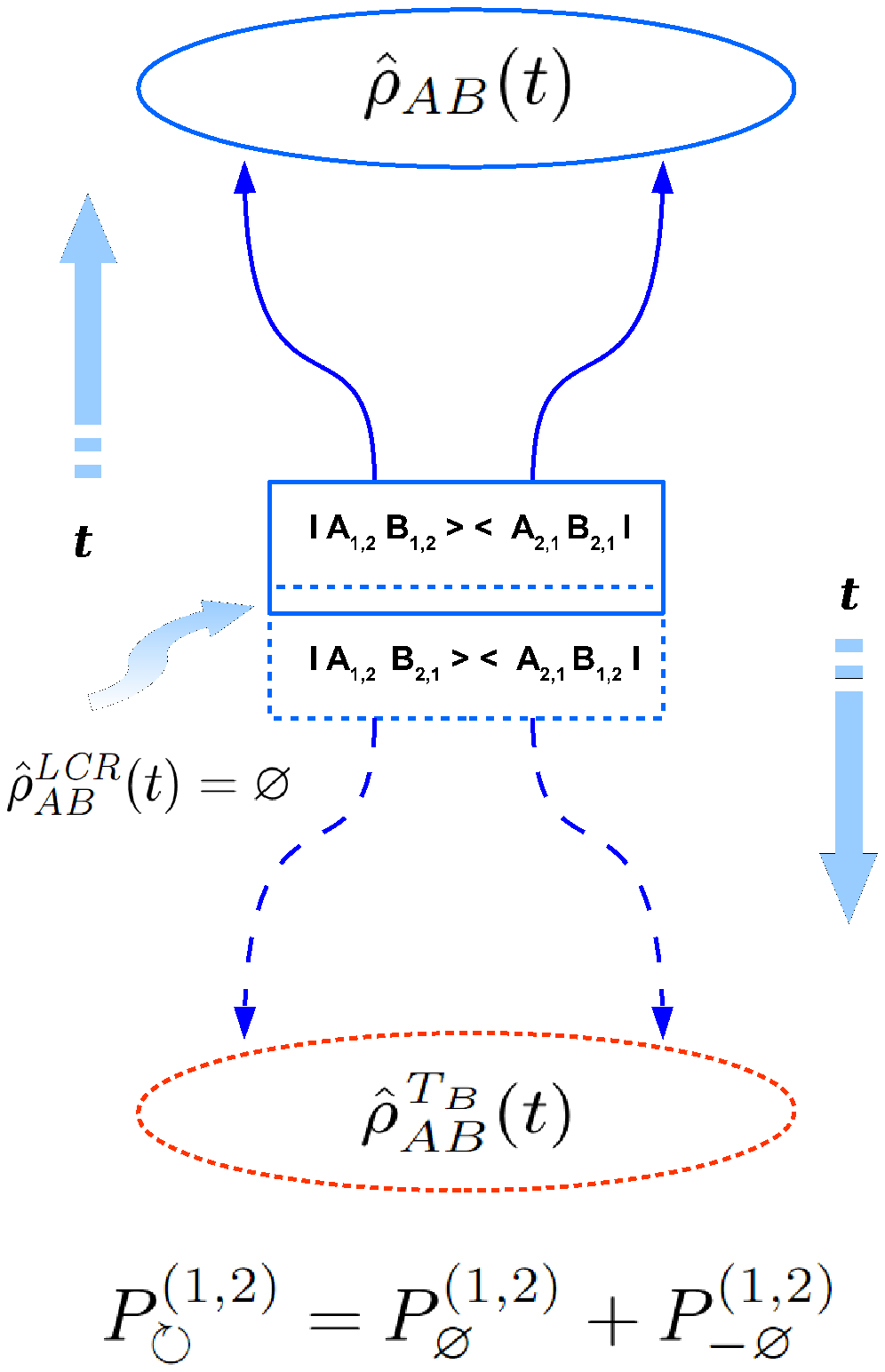}
\caption{Schematic picture of $ 4 \times 4 $ (two qubit- ) density matrix evolution in the \textit{entangled case} with respect to its partial transposition (pt-) procedure performed on $ B $ -subsystem taking into account two possible directions of a global time arrow in the entangled case. Square blocks are supposed to mark the moment of a given density matrix preparation at $ t=t_{LCR}=0 $. It is shown that in the entangled case the pt- transformed density matrix being defined  in the  $ 2 \otimes 2  $ Hilbert space, involves virtual quantum transitions (projectors in the dashed block) at the moment of its preparation. The latter projectors differ from those (projectors in the solid block) being responsible for the preparation of the original entangled density matrix as a whole. Hence, the partial transpose (pt-) transformation of the entire density matrix in one subsystem's subspace automatically changes the way of quantum states preparation in the another subsystem(s) being entangled with the former. Since \textit{no local causality reverse is possible in the entangled case} in the entangled case one has $ \hat{\rho}^{\textit{LCR}}_{AB}(t) =\varnothing $. }
\end{figure}

\begin{figure}
\includegraphics[height=13 cm,width=9 cm]{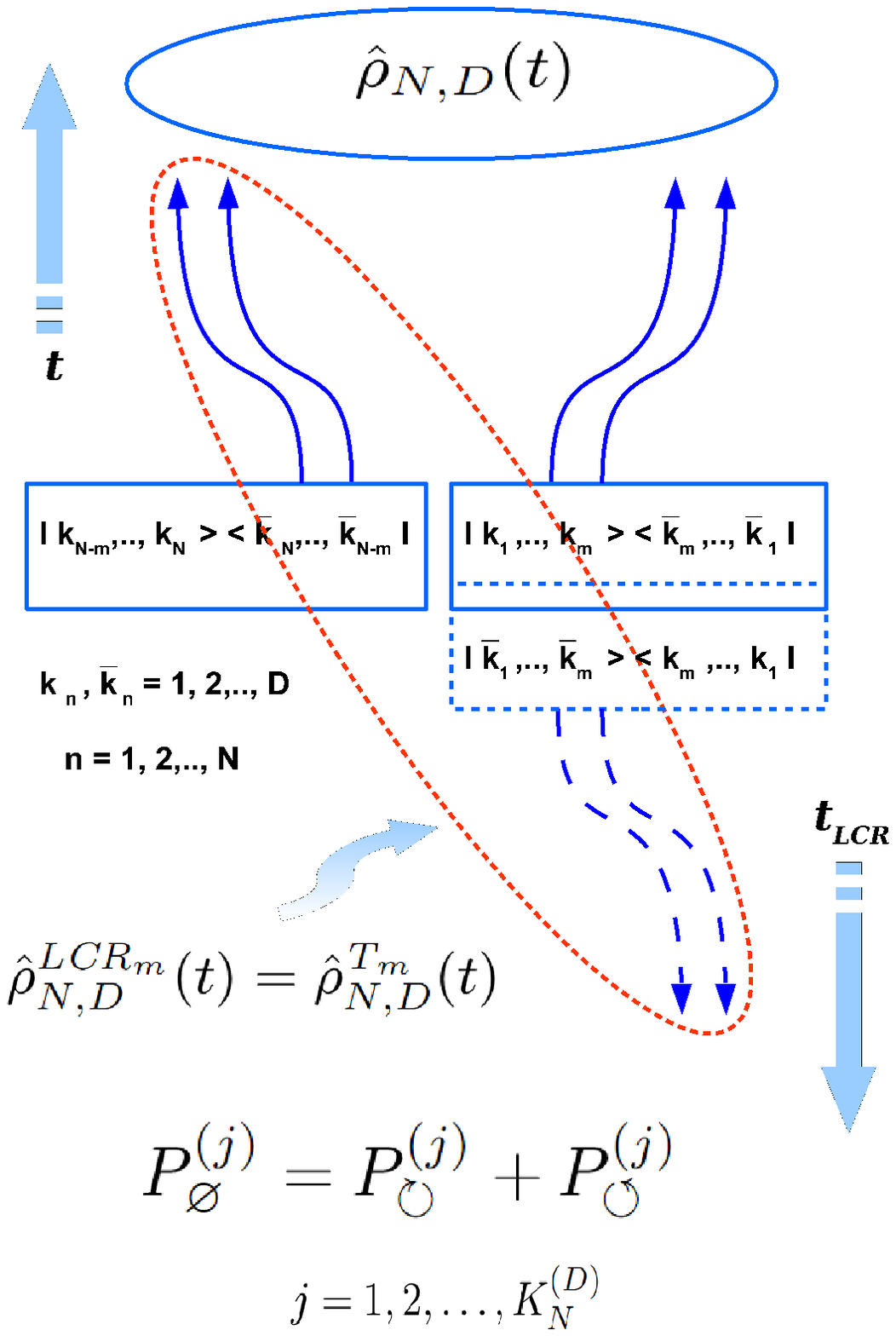}
\caption{Schematic picture of the evolution of density matrix defined in $ D^{\otimes N} $ Hilbert space (encoding $ N $ quantum subsystems of $ D $ eigenstates each) in the \textit{separable case} with respect to the \textit{local causality reversal} (LCR-) procedure performed on $ m $ among $ N $ its subsystems ($ m = 1,2,\ldots N-1 $). All the scheme represents just a mapping of the Fig.1 on the most general case where $ k=1,2,\ldots D $ (instead of $ 1,2 $ on Fig.1) and $ n=1,2,\ldots N $ ((instead of $ 1,2 $ on Fig.1), while $ m $ can be $  1,2,\ldots N-1 $ (instead of $ m=1 $ case on Fig.1). Besides that the meaning of all constituent blocks, arrows, etc. here is the same as for ones on Fig.1. As well, the entire explanation of all the separable LCR scheme is qualitatively the same as one for Fig.1. }
\end{figure} 

\begin{figure}
\includegraphics[height=13 cm,width=9 cm]{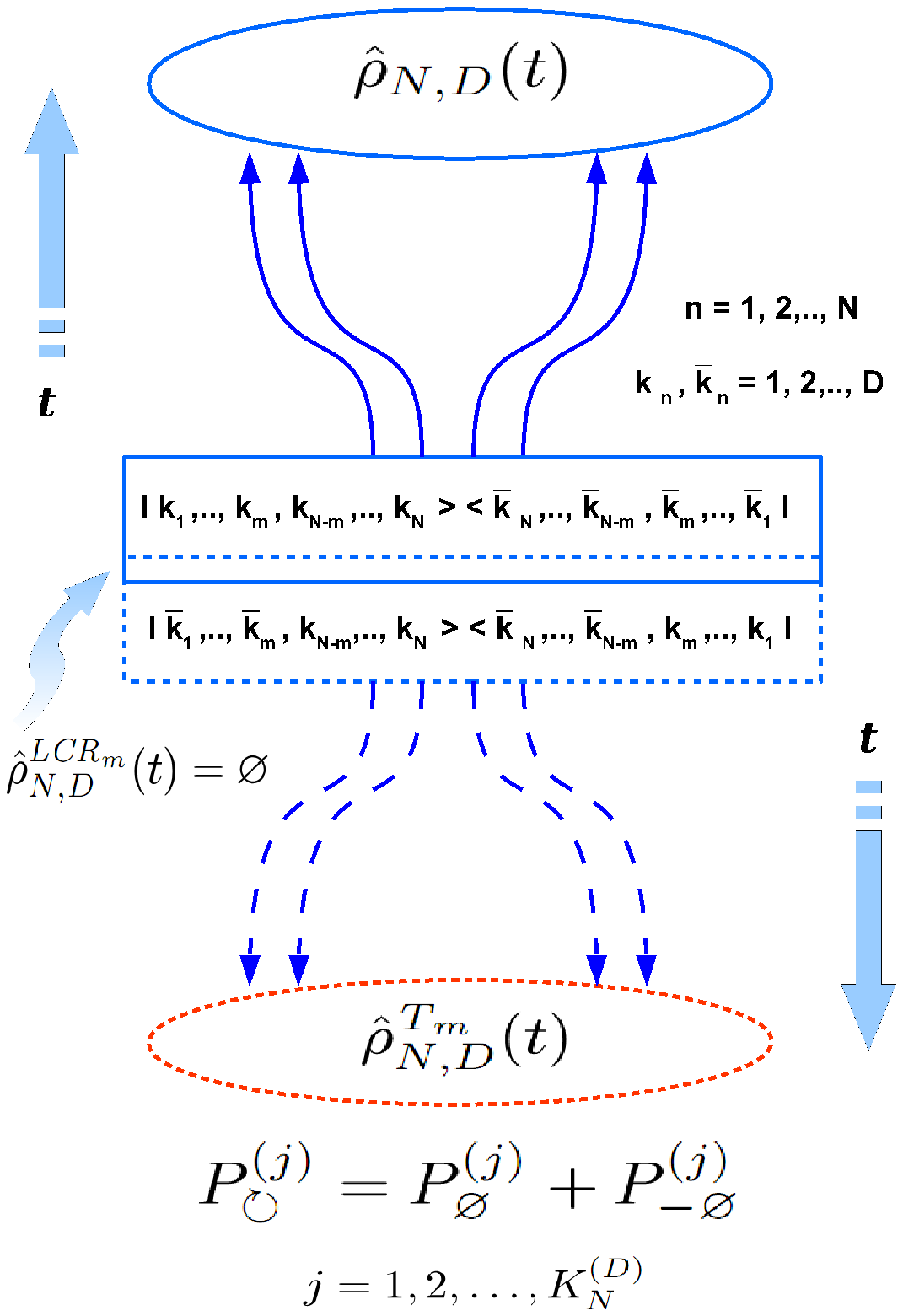}
\caption{Schematic picture of the evolution of density matrix defined in $ D^{\otimes N} $ Hilbert space (encoding $ N $ quantum subsystems of $ D $ eigenstates each) in the \textit{entangled case} with respect to two possible directions of a global time arrow and pt-transformation performed on $ m $ among $ N $ its constituent subsystems ($ m = 1,2,\ldots N-1 $). All the scheme represents just a mapping of the Fig.2 on the most general case where $ k=1,2,\ldots D $ (instead of $ 1,2 $ on Fig.2) and $ n=1,2,\ldots N $ (instead of $ 1,2 $ on Fig.2), while $ m $ can be $  1,2,\ldots N-1 $ (instead of $ m=1 $ case on Fig.2). Besides that the meaning of all constituent blocks, arrows, etc. here is the same as for ones on Fig.2. The  explanations behind all the entangled pt-scheme are the same as ones for Fig.2. }
\end{figure} 

Here one can see that equations (8,9) are just another mathematical form of Eqs.(2,3) in the separable and entangled case correspondingly. This means Eqs.(8,9) both represent an alternative form of the ppt-separability criterion, the form which remarkably has new general physical meaning. Especially, one can see that Eq.(8) is invariant under the lower symbols interchange $ \circlearrowright \longleftrightarrow \circlearrowleft $, while Eq.(9) is invariant under another lower symbols interchange $ \varnothing \longleftrightarrow - \varnothing $. Physically, this encodes two following statements: i) \textit{the probability of observer's ignorance about the quantum state of separable constituent subsystems of quantum system during its preparation is invariant with respect to the local causality reversal (LCR-) procedure performed on any separable subsystem of a given quantum system} and ii) \textit{the probability of a virtual quantum transition between two completely orthogonal compositions of base states of the entangled quantum system - remains invariant with respect to a reversal of a global time arrow}. Obviously, these two statements express physical meaning of Eqs.(8,9) and thus reveal a true physical content of the ppt-separability criterion for density matrices. From Eqs.(6-9) being valid for $ 2 \otimes 2 $ case it follows that in such the case

\begin{align}
\begin{split}  
\ \ W^{j}=\left\{ 
 \begin{matrix}
P^{(j)}_{\circlearrowleft}, \ \ for \ W^{j} \geq 0 \ \ (separable \ case) \ \\  
\\ 
 - P^{(j)}_{- \varnothing},  \ \ for \ W^{j} < 0 \ \  (entangled \ case) \
        \end{matrix} \right.
 \end{split}
\end{align}

Therefore, equations (8-10) in combination with definitions (4,5) uncover a true physical background for the determinants (2,3) in the most simple case of  $ 4 \times 4 $ density matrices for generic bipartite quantum system.

The remarkable practical advantage of the new separability criterion in the form of Eqs.(8-10) is that these causal relations (8,9) are very general and, basically, these relations are independent on the rank of underlying density matrix.  The latter fact has far-reaching consequences which go far beyond the $ 2 \otimes 2 $ case relations (8-10) have been derived for. Namely, Eq.(8) can be used as the definition of the probability $ P^{(j)}_{\circlearrowleft} $  in the separable case via two known quantities $ P^{(j)}_{\varnothing} $ and  $ P^{(j)}_{\circlearrowright} $ , whereas Eq.(9) can serve as the definition of the probability $ P^{(j)}_{- \varnothing} $ being specific for the entangled case via  known quantities $ P^{(j)}_{\varnothing} $ and  $ P^{(j)}_{\circlearrowright} $. In what follows we will see that the definitions of the probabilities $ P^{(j)}_{\varnothing} $ and  $ P^{(j)}_{\circlearrowright} $ both can be extended to the generic case of  $  D^{N} \times D^{N}$ -dimensional density matrices which describe $ N $-partite quantum systems consisting of $ N $ quantum subsystems with $ D $ eigenstates each . 

Thus, a central assumption of this paper is that \textit{the ignorance probability decomposition of Eq.(8) for separable density matrices and virtual transition probability decomposition of Eq.(9) for the entangled density matrices - both should be valid for generic density matrix defined in the $  D^{\otimes N} $ Hilbert space with arbitrary $ D $ and $ N $ describing quantum system of $ N $ quantum subsystems living in $ D $-dimensional Hilbert subspace each}. This, in turn, allows one to formulate in what follows a brand new universal causal separability/entanglement criterion for arbitrary $ D^{N} $ - dimensional density matrix without the necessity to calculate all its eigenvalues explicitly. 

\section{Generalisation of basic causal relations on the arbitrary $ D^{ \otimes N} $ case} 

One can see that basic Eqs.(8,9) which connect the fact of quantum states separability or entanglement with the causal decompositions of "ignorance"- or virtual quantum transition probabilities are very general since they do not contain any restrictions on the dimensionality of underlying density matrix of a composite quantum system. That is why, using the same logic which has led us to Eqs.(4-10) one can easily generalize Eqs.(4-10) and hence the ppt-separability criterion of Eq.(1) on a quite general case of arbitrary $ D^{N} $-dimensional density matrix which describes by its definition the ensemble of $ N $ quantum subsystems with $ D $ eigenstates of observable being measured in the case of arbitrary degree of entanglement between such quantum subsystems. 

Indeed, if each among $ N $ quantum subsystems  has $ D $ orthonormal eigenstates corresponded to $ D $ eigenvalues $ k^{(i)}_{n} $ ($ i=1,2,..,D $; $ n = 1,2,.., N $) of a certain observable $ k $ being measured for each subsystem then each among these states (which might be detected in the experiment manipulating given density matrix) can be written in the form $ \vert k^{(j)}_{n} \rangle $. Here index $ j=1,2,..,K^{(D)}_{N} $ denotes each of $ K^{(D)}_{N} $  \textit{distinct configurations} of the entire quantum system this particular subsystem's state is a part of. On how to calculate $ K^{(D)}_{N} $ one can see below. As well, index  $ k^{(j)}_{n} = 1,2,..,D $ marks orthonormal eigenstates of certain observable $ k $ which can be measured for each $ n $-th subsystem as a part of $ j $-th configuration of the entire ensemble - in what follows we will call such eigenstates simply as \textit{subsystem's base states}). Notice, that in the latter definition we have replaced concrete eigenvalues by the values of their indices $ k_{n} $ because in our subsequent generalized description  concrete observable eigenvalues are irrelevant.  Therefore, it is quite natural to define an arbitrary base state vector of the entire $ D^{ \otimes N} $ quantum system as its ($ j $-th) configuration  

\begin{eqnarray} 
\vert \lbrace k_{N} \rbrace^{(j)} \rangle =\prod^{N}_{n=1} \otimes \vert k^{(j)}_{n} \rangle  \\ \nonumber
\\ \nonumber
k = 1,..,D ; \ j = 1,.., K^{(D)}_{N}.  
\end{eqnarray}

Now in order to make our subsequent analysis of density matrices as general as possible we should pay some attention to the fact of either presence or absence of \textit{interactions} of any kind between $ N $ constituent quantum subsystems of the ensemble encoded by means of a given density matrix. Especially, here one has only two distinct types of situation: i) \textit{interaction between all subsystems of the ensemble "dies off" till the moment when the eigenvalues of the characteristic observable $ k $ for the ensemble are measured} - \textit{"non-interacting"} case; and ii) \textit{interaction between all subsystems of the ensemble still exists all the time including the moment when one measures the eigenvalues of certain characteristic observable $ k $ in all subsystems of the ensemble} - this one may think of as the \textit{"interacting"} case. Here the situation of the first type is one being mostly studied in the context of separability criteria, quantum state purification, etc. while the study of the separability in the second interacting case would shed some extra light on the problem of entanglement in a wide range of interacting quantum many-body systems of different nature. In what follows a subsequent general theory covers both these opposite cases.

\subsection{Configurations with no interaction between the parties at the moment of system measurement }
 
In this case of  density matrices describing $ N $ non-interacting (but possibly entangled with each other due to interaction between them in the past) quantum subsystems of dimensionality $ D \geq 2 $ each one can introduce a \textit{completely orthogonal configuration} of the state vectors of subsystems with respect to its arbitrarily chosen $ j $-th configuration  as the base vector of state of the entire system being completely orthogonal to a vector of state for chosen $ j $-th system configuration $ \vert \lbrace k_{N} \rbrace^{(j)} \rangle $. This yields 

\begin{eqnarray} 
\vert \lbrace \bar{k}_{N} \rbrace^{(l)} \rangle =  \prod^{N}_{n=1} \otimes \vert k'^{(l)}_{n}  \rangle_{(k \neq k'),(j \neq l)}  \\ \nonumber 
\\ \nonumber
 k,k' = 1,..,D ; \ j = 1,.., K^{(D)}_{N}; \ l=1,..,\bar{K}^{(D-1)}_{N}   
\end{eqnarray}

From Eq.(12) one has evident definition of the complete orthogonality between two  product states $ \vert \lbrace \bar{k}_{N} \rbrace^{(j)} \rangle $ and $ \vert \lbrace k_{N} \rbrace^{(j)} \rangle $ (and as well, for all  base states of constituent subsystems)  in each $ j $-th configuration of the entire system 

\begin{eqnarray} 
 \langle \lbrace k_{N} \rbrace^{(j)} \vert \lbrace \bar{k}_{N} \rbrace^{(j)} \rangle = \langle  k^{(j)}_{n} \vert  k'^{(j)}_{n} \rangle_{(k \neq k')} = 0  \\  \nonumber 
 \\ \nonumber
 \langle \lbrace k_{N} \rbrace^{(j)} \vert \lbrace k_{N} \rbrace^{(j)} \rangle = \langle  k^{(j)}_{n} \vert  k^{(j)}_{n} \rangle = 1  \\  \nonumber  
\\ \nonumber 
k,k' = 1,..,D ; \ j = 1,..,K^{(D)}_{N}.  
\end{eqnarray}

In Eqs.(11-13) index $ j $ runs all $ K^{(D)}_{N} $ \textit{distinct} configurations of base states of the entire quantum system, where the term "distinct" refers to all those configurations of the system which are neither identical nor completely orthogonal (in the sense of Eq.(13)) to each other. Since each of these configurations has $ (D-1)^{N} $ its completely orthogonal counterparts and since the total number of \textit{all} configurations of the system is simply $ D^{N} $ one can calculate a number of all distinct configurations denoting $ x=K^{(D)}_{N} $ from the simple relation $ x + x (D-1)^{N}= D^{N}$. Therefore, one has for the total number of all distinct configurations

\begin{eqnarray} 
 K^{(D)}_{N} = \left \lbrace \frac{1}{ \left[ \frac{1}{D^{N}} + \left( 1 - \frac{1}{D}\right)^{N}\right]} \right \rbrace_{in}  
\end{eqnarray}

and

\begin{eqnarray} 
\bar{K}^{(D)}_{N} = \left \lbrace \frac{(D - 1)^{N}}{\left[ \frac{1}{D^{N}} + \left( 1 - \frac{1}{D}\right)^{N}\right]} \right \rbrace_{\bar{in}}     
\end{eqnarray}

for the total number of system's configurations each being \textit{completely (or mutually)} orthogonal to one among $ K^{(D)}_{N} $ its distinct configurations. In Eqs.(14,15) symbols $ \lbrace \ldots \rbrace_{in} $ and $ \lbrace \ldots \rbrace_{\bar{in}} $ mean integer part of the number taken in such a way in order to provide the validity of the equality $ K^{(D)}_{N} + \bar{K}^{(D)}_{N} = D^{N} $ which should always be a positive integer number.  Notice, that any two distinct system's configurations $ \lbrace k_{N} \rbrace^{(j)} $ and $ \lbrace k_{N} \rbrace^{(j')} $ for two arbitrary $ j \neq j' $ are not completely orthogonal to each other. The latter (complete orthogonality) is the case only for the pairs  $ \vert \lbrace k_{N} \rbrace^{(j')} \rangle = \vert \lbrace \bar{k}_{N} \rbrace^{(j)} \rangle $ and $ \vert \lbrace k_{N} \rbrace^{(j)} \rangle $ in accordance with Eq.(13). 

Remarkably, one can extract some important information about possible collective quantum states of the entire $ D^{N} $-dimensional quantum system already from Eqs.(14,15) for several most common particular cases. First of all, since the cases $ D=1 $ and $ N=1 $ are trivial, everywhere in Eqs.(14,15) we put $ D \geq 2 $ and $ N \geq 2 $ though these formulas remain correct also for  $ D=1 $ (and arbitrary $ N \geq 1 $)- in this trivial case according to Eqs.(14,15) one has only one configuration with no completely orthogonal counterparts. One can see that for $ N=2 $ and $ D=2 $ (i.e. for two qubits, say, for two $ 1/2 $ spins) there exist two distinct configurations of the system with two completely orthogonal counterparts as it is prescribed by Eqs.(2). As well, for  $ N=3 $ and $ D=2 $ (i.e. for the system of three qubits, say,  three $ 1/2 $ spins) one has four distinct configurations and four configurations being completely orthogonal to the former. Increasing $ N $ further for $ D=2 $ case from Eqs.(14,15 )one obtains that the number of distinct configurations grows as $ 2^{N-1} $ having the same number of completely orthogonal "partners" . Therefore, in such the case one can subdivide all possible quantum states of the entire $ 2^{N} $-dimensional quantum system on just two subsets of system's configurations where each configuration from one subset has its mutually completely orthogonal counterpart in another subset. This agrees with e.g. a common EPR situation where $ D=2 $ orthogonal quantum states become shared among $ N $ parties or with all possible quantum states of $ 1/2 $ spin chain of $ N $ sites: all quantum states of this spin chain one can subdivide on the states of just two spin sub-lattices where each sub-lattice has its spin configurations being completely orthogonal to corresponding spin configurations from another spin sub-lattice. 

On the other hand, as it can be seen from Eqs.(14,15), completely different situation takes place in the limit $ D \rightarrow \infty $ ( for arbitrary $ N \geq 2 $ ), say, for density matrix which describes two or more quantum oscillators. In this case one has asymptotically  $ K^{(D)}_{N} \rightarrow 1 $ and  $ \bar{K}^{(D)}_{N} \rightarrow \infty $. It means that \textit{in the limit $ D \rightarrow \infty $ at arbitrary $ N \geq 2 $ one has no distinct (i.e. not completely orthogonal) configurations of the entire non-interacting quantum ensemble, but, instead, all available configurations of subsystems' eigenstates are completely orthogonal to each other}. In other words, Eqs.(14,15) prove the fact that \textit{two or more identical quantum oscillators non-interacting with each other can never be prepared and then measured in the same quantum state simultaneously}. And this is nothing more than a manifestation of a famous \textit{no-cloning theorem} from quantum mechanics.

In general, if one takes certain fixed configuration of the ensemble of $ N $ non-interacting (at the moment of measurement) $ D $ -dimensional quantum subsystems one will count $ N_{o.c.}=(D-1)^{N} $ another configurations being completely orthogonal to the former, while the number of all available virtual quantum transitions $ N_{v.t.} $ given configuration may experience simultaneously will be only $ D-1 $ in this non-interacting case.  

\subsection{Configurations in the interacting ensemble of $ N $ quantum systems being prepared and measured}

The case of $ N $ interacting $ D $-dimensional quantum systems forming given density matrix is more subtle. First of all, the presence of interaction between \textit{all} $ N $ constituent subsystems of given quantum ensemble means that generally speaking one has  $ \langle  k^{(j)}_{n} \vert  k^{(j)}_{n'} \rangle_{n \neq n'} \neq 0 $ , in contrast with Eq.(13) for \textit{all} scalar products between base state vectors which \textit{would be} orthogonal eigenvectors of a measurement apparatus for observable $ k $  referred to two different ($ n $-th and $ n' $-th) independent subsystems of the ensemble in the case if these two measured subsystems were disconnected from each other. But since all subsystems now are coupled (or equivalently, interacting) at the moment of observable measurement (as well as at the moment of system preparation) the base state vectors of different subsystems now all become not independent from each other (or equally - the latter become non-orthogonal to each other) due to interaction. And this represents a very general type of quantum many-body system where quantum transitions between states of different subsystems become possible  due to overlap (or tunnelling) between quantum states of distinct quantum subsystems of  the ensemble. Thus, if the interaction involves all $ N $ subsystems of the ensemble then corresponding orthonormal basis in Hilbert space shrinks to only $ D $ orthogonal eigenvectors for any number $ N $ of interacting subsystems (here $ D $ is the dimensionality of a Hilbert subspace for each subsystem of the ensemble). 

Of course, following the analogy with previous non-interacting case, one can still introduce the same state vectors of system configurations as ones from Eqs.(11,12) but, obviously, in the interacting case the quantity $ K^{(D)}_{N}= D^{N-1} $ and $ \bar{K}^{(D)}_{N}= (D-1)D^{N-1} $ both will differ from formulas (14,15) for the non-interacting case. At the same time, as it becomes clear from the above, if one takes certain fixed configuration of interacting ensemble of $ N $  $ D $-dimensional quantum subsystems then there will be only $ N_{o.c.}=N^{(I)}_{o.c.}=(D-1)$ its \textit{completely orthogonal} counterparts (in the sense of Eq.(13)) but for the maximal number of possible virtual quantum transitions $ N_{v.t.} $ between given configuration and all its completely orthogonal "partners" in the "true" orthonormal basis of $ \vert \lbrace \bar{k}_{N} \rbrace^{(j)} \rangle $ state vectors one will have $ N_{v.t.}= N^{(I)}_{v.t.} = (D-1)^{N} $ in contrary with the previous non-interacting case. Basically, different values of quantities $ N_{o.c.}$ and $ N_{v.t.} $ are the only distinctions between the cases of "interacting" and "non-interacting" ensembles in the above introduced description in terms of ensemble configurations. Therefore, below everywhere in our description we will concern both cases of  "interacting" and "non-interacting" ensembles of $ N $ $ D $-dimensional quantum subsystems.

\section{Density matrices and universal causal separability criterion in terms of configurations}

Obviously, in the most general case (which includes both interacting and non-interacting situations) in the basis of orthonormal state vectors of Eq.(11) (or equally, in the basis of \textit{all system's configurations}) arbitrary density matrix $ \hat{\rho}^{(D)}_{N} $ defined in the $ D^{ \otimes N} $ Hilbert space takes following form

\begin{eqnarray} 
\hat{\rho}^{(D)}_{N}=\sum_{j,j'=1}^{D^{N}} \vert \lbrace k_{N} \rbrace^{(j)} \rangle \rho_{\lbrace k \rbrace_{j} \vert \lbrace k \rbrace_{j'}} \langle \lbrace k_{N} \rbrace^{(j')} \vert 
\end{eqnarray}

here each among summation indices $ j,j' $ runs all $ D^{N} $ possible configurations of the entire quantum system with respect to the quantum states of its  $ N $ constituent subsystems and according to Eqs.(11-15) all these configurations can be divided on two classes: distinct and completely orthogonal (to distinct) ones. 

Remarkably, density matrix in its most general "configuration" representation (16) already contains all its matrix elements in the form $ \rho_{\lbrace k \rbrace_{j} \vert \lbrace k \rbrace_{j'}} $ i.e. in the form which is already familiar to us from Eqs.(2-5) for the simplest $ 2 \times 2 $ case. This fact allows one to generalize all probabilistic definitions (4,5) as well as all the subsequent relations (6-10) being derived for the simplest bipartite quantum system in the $ 2 \otimes 2 $ Hilbert space on the most general $ D^{\otimes N} $ case. 

Especially, for the "ignorance probability" $ P^{(j)}_{\varnothing} $, which reflects the uncertainty in observer's knowledge about the realization of either $ \lbrace k_{N} \rbrace^{(j)} $ configuration or one among its completely orthogonal $ \lbrace \tilde{k}_{N} \rbrace^{(j)}  $ counterparts in given N-partite quantum system from $ D^{\otimes N} $ Hilbert space, one can write

\begin{eqnarray} 
P^{(j)}_{\varnothing}=\rho_{\lbrace k \rbrace_{j} \vert \lbrace k \rbrace_{j}} \left(  \sum_{l=1}^{N_{o.c.}} \rho_{\lbrace \bar{k} \rbrace_{l} \vert \lbrace \bar{k} \rbrace_{l}} \right). 
\end{eqnarray}

where $ N_{o.c.} $ is the number of completely orthogonal configurations $ \lbrace \bar{k}_{N} \rbrace^{(l)} $ for given configuration $ \lbrace k_{N} \rbrace^{(j)} $ chosen, $ N_{o.c.} $ is different in the cases of non-interacting and interacting subsystems in the ensemble. The difference between general definition (17) and its $ 2\otimes 2 $ particular case of Eq.(4) is due to the fact that for each $ j $-th configuration of the system  there exist $ N_{o.c.} $ completely orthogonal configurations $ \lbrace \bar{k} \rbrace_{l} $ one should sum over in order to take them all into account in the "ignorance" probability. 

Analogously, the generalisation of definition (5) for the probability $ P^{(j)}_{\circlearrowright} $  of a virtual quantum transition of the entire quantum system from $ D^{\otimes N} $ Hilbert space between its completely orthogonal configurations  $ \lbrace k \rbrace_{j} $ and $ \lbrace \bar{k} \rbrace_{l} $ takes the form

\begin{eqnarray} 
P^{(j)}_{\circlearrowright(m)} = \sum_{l=1}^{N_{v.t.}}\rho^{T_{m}}_{\lbrace k \rbrace_{j} \vert \lbrace \bar{k} \rbrace_{l}} \rho^{T_{m}}_{ \lbrace \bar{k} \rbrace_{l} \vert \lbrace k \rbrace_{j}} 
\end{eqnarray}

where the symbol $ \rho^{T_{m}} $ means operation of partial transposition of a density matrix $ \hat{\rho} $ simultaneously in the $ m $ ( $ m \leq (N -1) $ ) subspaces of its Hilbert space (the latter procedure corresponds either to LCR-operation performed on $ m $ subsystems in the separable case (when the entire subset of $ m $ subsystems is separable from the rest of subsystems) or to the global causality/time arrow reversal - in the entangled case, when the subset of $ m $ subsystems as a whole is entangled with the rest of system. In this definition the summation is due to the fact that for each constituent state  $\vert k^{(j)}_{n} \rangle $ of each subsystem in the $ j $-th distinct configuration of the system the $ N_{v.t.} $ virtual transitions to respective completely orthogonal configurations are possible. Therefore, each $ j $-th configuration of the system can transfer virtually only to one among  $ N_{v.t.}  $ its another completely orthogonal configurations. As it has been already mentioned in the above, the numbers $ N_{v.t.}  $ and $ N_{v.t.}  $ are different for the cases of non-interacting and interacting constituent subsystems in the ensemble.

As it follows from all the above, two general \textit{causal-invariant} relations of Eq.(8) and Eq.(9) for separable and entangled density matrices, correspondingly, together with definition (10) remain valid for \textit{any} density matrix (16) defined in $ D^{ \otimes N} $ Hilbert space with arbitrary $ D $ and $ N $. Therefore, new definitions (17,18) for the ignorance- and virtual quantum transition probabilities for arbitrary density matrices of the rank $ \leq D^{N} $ allow one to generalize the "causal" form of the separability/entanglement criterion of Eqs.(3,10) on the case of arbitrary $ D^{\otimes N} $ Hilbert space in a following way 

\begin{eqnarray} 
\nonumber
W^{(j)}_{m}= \left( \rho_{\lbrace k \rbrace_{j} \vert \lbrace k \rbrace_{j}} \sum_{l=1}^{N_{o.c.}} \rho_{\lbrace \bar{k} \rbrace_{l} \vert \lbrace \bar{k} \rbrace_{l}} \right)  \\ \nonumber
 - \left( \sum_{l=1}^{N_{v.t.}}\rho^{T_{m}}_{\lbrace k \rbrace_{j} \vert \lbrace \bar{k} \rbrace_{l}} \rho^{T_{m}}_{ \lbrace \bar{k} \rbrace_{l} \vert \lbrace k \rbrace_{j}} \right)  \\
\\ \nonumber
 k=1,2,..,D;  \ \ j = 1,2,.., K^{(D)}_{N}, 
\end{eqnarray}

where $ K^{(D)}_{N} $ is a number of all distinct (i.e. not completely orthogonal) configurations of the ensemble of $ N $  $ D $-dimensional quantum subsystems joined into one either interacting (coupled) or not quantum ensemble with

\begin{align}
\begin{split}  
 \ \left\{ 
 \begin{matrix}
 N_{o.c.} = (D - 1)^{N} , \ \ N-free \ ensemble   \\  
\\ 
 N_{o.c.} = (D - 1) ,  \ \ N-coupled \ ensemble   \\ 
        \end{matrix} \right.
 \end{split}
\end{align}

and 

\begin{align}
\begin{split}  
\ \left\{ 
 \begin{matrix}
 N_{v.t.} = (D - 1), \ \  N-free \ ensemble   \\  
\\ 
 N_{v.t.} = (D - 1)^{N}, \ \  N-coupled \ ensemble   \\
        \end{matrix} \right.
 \end{split}
\end{align}

here terms $ N-coupled $ (and $ N-free $) mark two opposite situations where all $ N $ quantum subsystems of the ensemble are  interacting (non-interacting) with each other at the moment of time when one measures matrix elements of a given density matrix of the ensemble.

Then, analogously to Eq.(10), for \textit{each}  $ j $-th distinct configuration $ \lbrace k \rbrace_{j} $ of $ N $ subsystems' base quantum states one has the criterion

\begin{align}
\begin{split}  
 W^{(j)}_{m}=\left\{ 
 \begin{matrix}
P^{(j)}_{\circlearrowleft(m)}, \ for \ W^{(j)} \geq 0, \\
\\
  \ if \ \lbrace k_{N} \rbrace^{(j)} \ is \ m-separable;    \\  
\\ 
 - P^{(j)}_{- \varnothing(m)},  \  for \ W^{(j)} < 0, \\
 \\
   \ if \ \lbrace k_{N} \rbrace^{(j)} \ is \ m-entangled;   
        \end{matrix} \right.
 \end{split}
\end{align}

For arbitrary density matrix defined in $ D^{ \otimes N} $ Hilbert space there exist $ K^{(D)}_{N} $ separability/entanglement conditions of the type (19,22) - one for each distinct configuration of subsystems' eigenstates. Some of these equations may appear to be equal to each other, thus, sufficiently reducing  the total number of restrictions system should obey to be separable/entangled. As the illustration let us consider the particular case of $ N $-free ensemble. Then, for example, in the case $ N=2 $ and $ D=2 $ Eqs.(19,22) reproduce Eqs.(2,10) of ppt (or PH-) separability criterion for $ 4 \times 4 $ density matrices. For $ N=3 $, $ D=2 $ one has four equations of the type (19,22) for all distinct configurations of the system and four their completely orthogonal counterparts. In the limit $ N \rightarrow \infty $ (or $ D \rightarrow \infty $) the number of conditions (19,22) for such quantum system also tends to infinity, however, this tendency can be compensated by the possible equivalence of a majority of such equations in some particular cases.  

Therefore, in the same way as it takes place for the ppt-separability criterion in $ 2 \otimes 2 $ case, one can claim that the quantum state of $ N $-partite quantum system described by arbitrary $ D^{N} $-dimensional density matrix is separable only if \textit{all} its distinct configurations of subsystems' eigenstates are separable and, otherwise, given $ D^{N} $-dimensional density matrix describes entangled $ N $-partite quantum system if \textit{at least one} among its distinct configurations is entangled according to Eq.(22). 

As it has been already mentioned in the above the entire causal concept which is summarized in two fundamental causal decompositions of Eqs.(8,9) for separable and entangled cases remains the same in general $ D^{\otimes N}  $ case as one for the case of $ 2 \otimes 2 $ Hilbert space. That is why formulas (8,9) and all the causal scheme depicted on Fig.1 for the separable- and on Fig.2 -for the entangled states of $ 4 $-dimensional bi-partite quantum system changes only a little for density matrices from $ D^{\otimes N}  $ Hilbert space. Corresponding schemes for the absolutely separable and for the entangled quantum states of generic $ N $-partite quantum system one can see on Fig.3 and Fig.4, correspondingly. Therefore, all qualitative explanations behind Figs.3,4 remain the same as ones for Figs.1,2.

Remarkably, since any density matrix (16) can be represented in the form

\begin{eqnarray} 
\hat{\rho}^{(D)}_{N}=\sum_{j=1}^{K^{(D)}_{N}} \vert \lbrace k_{N} \rbrace^{(j)} \rangle \rho_{\lbrace k \rbrace_{j} \vert \lbrace k \rbrace_{j}} \langle \lbrace k_{N} \rbrace^{(j)} \vert \\ \nonumber 
+ \sum_{j=1}^{\bar{K}^{(D)}_{N}} \vert \lbrace \bar{k}_{N} \rbrace^{(j)} \rangle \rho_{\lbrace \bar{k} \rbrace_{j} \vert \lbrace \bar{k} \rbrace_{j}} \langle \lbrace \bar{k}_{N} \rbrace^{(j)} \vert \\ \nonumber
+ \sum_{j=1}^{K^{(D)}_{N}} \sum_{j'=1}^{\bar{K}^{(D)}_{N}} \vert \lbrace \bar{k}_{N} \rbrace^{(j')} \rangle \rho_{\lbrace \bar{k} \rbrace_{j'} \vert \lbrace k \rbrace_{j}} \langle \lbrace k_{N} \rbrace^{(j)} \vert \\ \nonumber 
+ \sum_{j=1}^{K^{(D)}_{N}}\sum_{j'=1}^{\bar{K}^{(D)}_{N}} \vert \lbrace k_{N} \rbrace^{(j)} \rangle \rho_{\lbrace k \rbrace_{j} \vert \lbrace \bar{k} \rbrace_{j'}} \langle \lbrace \bar{k}_{N} \rbrace^{(j')} \vert \\ \nonumber
+ \ \lbrace \ the \ rest \ of \ matrix \ elements 
\  \rbrace 
\end{eqnarray}

one can see that causal separability/entanglement criterion (or simply \textit{CS-criterion}) of Eqs.(19,22) is  \textit{straightforward} involving only four types of density matrix elements (i.e. ones from the first four lines in the r.h.s. of Eq.(23) ) and, importantly, \textit{this novel CS- criterion does not require any explicit calculation of eigenvalues or any other complicated manipulations with matrix elements of arbitrary $ D^{N} $-dimensional density matrix of interest}. 

At this point, one clearly sees that Eqs.(19-22) together with Eqs.(14,15) represent a completely novel heuristic type of separability/entanglement criteria of causal nature (8,9) - called the \textit{CS - criterion} for arbitrary $ D^{N} $-dimensional density matrices from $ D^{\otimes N}  $ Hilbert spaces describing initially prepared ensemble of $ N $ quantum subsystems being either separable or entangled, interacting or not with each other, where each subsystem is defined in its own $ D $-dimensional Hilbert subspace. 

General relations of Eqs.(8-10), Eqs.(14,15) together with new causal separability/entanglement criterion of Eqs.(17-22) all represent main findings of this paper and will be illustrated on several most important examples in what follows. 

\section{ Discussion}

To illustrate the power of a novel causal separability (CS-) criterion derived in the above let us consider probably its most important generic example: one-parametric family of density matrices $ \hat{\rho}^{(D)}_{N}(p) $ those defined in the $ D^{\otimes N}  $ dimensional Hilbert space (with arbitrary number of  subsystems $ N \geq 2  $ and arbitrary  $ D \geq 2 $ dimensionality of  each subsystem's Hilbert subspace) and all being only one- ($ p $-) parameter-dependent. In what follows it will be reasonable to call all density matrices of such type as \textit{equally connected} ( EC-) ones or, alternatively, \textit{Werner connected}  since in its  $ D \otimes D  $  realization such matrices encode Werner states of two-partite quantum system where each among two parties (or subsystems) is defined in $ D $-dimensional Hilbert space. Especially, for the particular case $ D=2 $ the latter  states of $ D \otimes D  $  quantum  system encode usual two-qubit Werner states, which turn out to be so-called \textit{Werner-entangled} see e.g. \cite{5,6,8} and some examples in below. In the literature on separability criteria for $ 2 \otimes 2 $ case people distinguish two classes of one-parametric EC-density matrices: a) one discriminating between one completely orthogonal configuration and the rest of configurations with equal transition rates between distinct pairs of completely orthogonal configurations ( Peres \cite{5} ) and b) EC-density matrices with discrimination between two distinct (not completely orthogonal) configurations and with strong discrimination between corresponding transition rates of these two types of distinct configurations to their completely orthogonal counterparts ( Horodecki \cite{6} ). Here we consider the $ D^{\otimes  N} $ generalisations of both these classes - called in below as a)- and b) ones - on the family of  $ D^{N} $-dimensional EC-density matrices. The details of the derivation of all subsequent formulas one can see in the Appendix A together with some additional comments on the validity of each sub-case being considered.


\subsection{Different realizations of the equally connected (EC-) one-parametric family of density matrices}

Let us emphasize the most important findings from the previous section concerning new causal separability and entanglement criterion for EC- family of density matrices acting in the $ D^{\otimes N}  $ Hilbert space. First of all, the term "equally (or Werner-) connected" means the fact that \textit{all $ N $ subsystems of a given quantum system or ensemble are the same having $ D $ dimensions in the Hilbert subspace for each subsystem} and  \textit{each subsystem of a given quantum system or ensemble is connected with the rest of the ensemble to the same degree}. Then a) and b) classes of equally- (EC-) or Werner-connected density matrices represent two basic possibilities for the off-diagonal matrix elements of one-parametric density matrices. 

To proceed further one needs to represent a generic ( -weakly mixing, non-interacting, see below) case of a-class of the EC-density matrix for the entire $ D^{ \otimes N} $ quantum system in its configuration representation of Eq.(23). As the result one yields (for the details of the derivation one can see the Subsection 1 of the Appendix A)

\begin{eqnarray} 
\nonumber
\hat{\rho}^{(a)}_{N}(p)= \\ \nonumber
\sum_{j=1}^{K^{(D)}_{N}} \vert \lbrace k_{N} \rbrace^{(j)} \rangle \frac{ (1 - \vert p \vert )^{m(j)}\vert p \vert^{N-m(j)}}{(D-1)^{N-m(j)}} \langle \lbrace k_{N} \rbrace^{(j)} \vert \\ \nonumber 
+ \sum_{j=1}^{\bar{K}^{(D)}_{N}} \vert \lbrace \bar{k}_{N} \rbrace^{(j)} \rangle \frac{(1 - \vert p \vert )^{N-m(j)}\vert p \vert^{m(j)}}{(D-1)^{m(j)}} \langle \lbrace \bar{k}_{N} \rbrace^{(j)} \vert \\ \nonumber
+ \sum_{j=1}^{K^{(D)}_{N}} \sum_{j'=1}^{\bar{K}^{(D)}_{N}} \vert \lbrace \bar{k}_{N} \rbrace^{(j')} \rangle \frac{ p^{N}}{(D-1)^{N}}  \langle \lbrace k_{N} \rbrace^{(j)} \vert \\ \nonumber 
+ \sum_{j=1}^{K^{(D)}_{N}}\sum_{j'=1}^{\bar{K}^{(D)}_{N}} \vert \lbrace k_{N} \rbrace^{(j)} \rangle \frac{ p^{\ast N}}{(D-1)^{N}}  \langle \lbrace \bar{k}_{N} \rbrace^{(j')} \vert \\ \nonumber
\\ \nonumber
\\
 m(j)= m =1,2,..,N.  \ \ \ \ 
\end{eqnarray}

In Eq.(24) the integer value $ m(j)= m = 1,2,..,N $ depends on the particular ($ j $-th) configuration of  $ N $ base quantum states associated with particular set of measured eigenvalues of observable $ k $  in all $ N $ subsystems of the ensemble. Obviously, because $ K^{(D)}_{N} \gg N $ and $ \bar{K}^{(D)}_{N} \gg N $, any a-class EC-density  matrix of the form (24) has equivalent matrix elements for each value of $ m(j)= m = 1,2,..,N $. Another important observation here is that for any two completely orthogonal configurations $ \lbrace k_{N} \rbrace^{(j)} $ and $ \lbrace \bar{k}_{N} \rbrace^{(j)} $ of subsystems' base states the corresponding power $ m_{j} $ in Eq.(24) should be the same.

In Eq.(24) the term a)-class describes equal off-diagonal matrix elements which encode equal probabilities of quantum transitions between two arbitrary completely orthogonal configurations of subsystems' orthogonal base states with respect to eigenvalues of certain local observable $ k $ in the entire  N-partite quantum system (or statistical ensemble). 

Another possibility for the off-diagonal matrix elements is realized in the b)-class of EC-density matrices where certain quantum transitions between completely orthogonal system's configurations are discriminated as compared to another ones. 

One can write down a generic (weakly mixing non-interacting, see below) b-class of EC-density matrix in the basis of its configurations  similarly to Eq.(24) and being pt-transformed in the subspaces of $ m $ parties of the entire ensemble as follows (the details of the derivation are represented in the Subsection 4 of the Appendix A)

\begin{eqnarray} 
\nonumber
\hat{\rho}^{(b)}_{N[m]}(p)= \\ \nonumber
\sum_{j=1}^{K^{(D)}_{N}} \vert \lbrace k_{N} \rbrace^{(j)} \rangle \frac{ (1 - p )^{m(j)} p^{N-m(j)}}{(D-1)^{N-m(j)}} \langle \lbrace k_{N} \rbrace^{(j)} \vert \\ \nonumber 
+ \sum_{j=1}^{\bar{K}^{(D)}_{N}} \vert \lbrace \bar{k}_{N} \rbrace^{(j)} \rangle \frac{(1 - p )^{m(j)} p^{N-m(j)}}{(D-1)^{m(j)}} \langle \lbrace \bar{k}_{N} \rbrace^{(j)} \vert \\ \nonumber
+ \sum_{j=1}^{K^{(D)}_{N}} \sum_{j'=1}^{\bar{K}^{(D)}_{N}} \vert \lbrace \bar{k}_{N} \rbrace^{(j')} \rangle \frac{ (1 - p)^{m(j)-m} p^{N-m(j)+ m}}{(D-1)^{N}}  \langle \lbrace k_{N} \rbrace^{(j)} \vert \\ \nonumber 
+ \sum_{j=1}^{K^{(D)}_{N}}\sum_{j'=1}^{\bar{K}^{(D)}_{N}} \vert \lbrace k_{N} \rbrace^{(j)} \rangle \frac{(1 - p)^{m(j)-m}p^{N-m(j)+ m}}{(D-1)^{N}}  \langle \lbrace \bar{k}_{N} \rbrace^{(j')} \vert \\ \nonumber
\\ \nonumber
\\
 m=\pm1,\pm2,..,\pm(N-1); \ m(j)=1,2,..,N.  \ \ \ \ 
\end{eqnarray}

In Eq.(25) the numbers $ K^{(D)}_{N} $ and $ \bar{K}^{(D)}_{N} $ of distinct and completely orthogonal configurations are from Eqs.(14,15) and the subscript $ [m] $ in the l.h.s. of equation as well as respective integer powers of $ \pm m $ in its r.h.s. are both the result of the pt-procedure applied to a given density matrix in the subspaces of $ m $ subsystems of the ensemble resulting in $ m $ non-equivalent permutations of $ (1-p) $ and $ p $ factors in the off-diagonal matrix elements. This fact is reflected in two sorts of the off-diagonal matrix elements in the entire $ D^{N}  $ dimensional density matrix being proportional to different powers of just two different numbers $ p $ and $ (1-p) $ which account for probabilities of different base states in each subsystem of the ensemble. Parameter $ p $ in Eqs.(24,25) is  the \textit{ entanglement parameter} of given EC-density matrix of Eq.(24) or Eq.(25), it accounts for such a procedure of density matrix preparation/evolution which took place in the system's past and, thus, which had caused a given degree of entanglement in the system. Obviously, $ p $ in the most general case is a certain complex function of time (see details in the Subsection 1 of the Appendix A), but in what follows without loss of generality one can assume $ p $ to be just a known real number being specific for the moment of time given density matrix is defined at. Quantum states of different EC-ensembles encoded by means of EC-density matrices of Eqs.(24,25) are often called in the literature as \textit{isotropic} quantum states \cite{28,29}. For some examples of concrete definitions and calculations of the entanglement parameters one can see e.g. Refs.\cite{27,34}.   

Obviously, the a)-class of EC-density matrices can be realized in those finite or infinite quantum statistical ensembles and quantum many-body systems, where all constituent subsystems are uniformly connected (i.e. uniformly mixed or uniformly interacting) with each other and where different eigenstates of certain chosen local observable $ k $ in the entire system all have equal probabilities  in the measurement outcome (except just one eigenstate which can be referred to as the most probable "ground state", or in contrary, the least probable "excited" state of each subsystem with the probability $ (1-p) $ depending on the value of $ p $). The latter situation can be realized when e.g. $ (D-1) $  base states from the chosen orthonormal basis in each subsystem of the ensemble are degenerate with respect to e.g. the energy of these configurations. This can take place e.g. in many examples of quantum spin chains, ensembles of weakly coupled qubits of different nature, etc. such as ones e.g. from Refs.\cite{11,12,16,17,18,19,20,21} . 

On the other hand, the b)-class of EC-density matrices describes all situations where \textit{several} base quantum states in the chosen $ D $-dimensional basis of local observable orthogonal eigenstates for each quantum system (or statistical ensemble) are more probable in observable measurement outcomes than all another eigenstates in each subsystem. Such the situation takes place in all those quantum many-body systems where at least two (or more) preferable base quantum states can be identified for each party of the system. The examples are represented by all those multi-partite quantum systems where several base quantum states for each party of the system are well-separated from the others (e.g. in bosonic systems including ensembles of coupled quantum oscillators, Rydberg atoms in photon cavities with well-separated ground and excited quantum states, etc. such as ones from Refs. \cite{2,4,13,14,15,23}) .

Further, the constituent parts (or subsystems) of a given N-partite equally connected (EC-) quantum system or of the statistical ensemble described by means of EC-density matrix of either a)- or b)-class can be subdivided on two additional subcategories with respect to the values of the off-diagonal matrix elements between the completely orthogonal configurations of the entire system. Those are: i) \textit{weakly mixing (w.)} EC-density matrices and ii) \textit{strongly mixing (s.) } EC-density matrices. The weakly mixing case is characterized by the uniform renormalisation of all such off-diagonal matrix elements by the factors of $ \frac{1}{(D-1)^{N}} $ and hence describes physical situation, where all constituent subsystems of a given quantum system (or ensemble) are only weakly coupled to each other. 

Therefore, substituting respective matrix elements from Eq.(24) into Eqs.(17-19) one obtains for arbitrary (non-interacting, i.e. N-free, see below)  $ D^{N} $ dimensional weakly mixing  a-class of EC-density matrix (see details in the Subsection 1 of the Appendix A)

\begin{eqnarray} 
W^{(a.w.)}(p)=  \vert p \vert^{N} \left\lbrace  (1 - \vert p \vert )^{N} -  \frac{\vert p \vert^{N}}{(D-1)^{2N-1}}  \right\rbrace. \nonumber \\   
\end{eqnarray} 

Now solving inequalities (22) with respect to the relation (26) one obtains following condition on the entanglement parameter $ \vert p \vert $  which determines separability or entanglement of arbitrary a-class N-free  weakly mixing  EC-density matrix from $ D^{\otimes N} $ Hilbert space 

\begin{align}
\begin{split}  
\left\{ 
 \begin{matrix}
  p_{th} < \vert p \vert \leq 1 , \  \ -entangled    \\  
\\ 
  0 \leq \vert p \vert \leq  p_{th}, \  \ -separable  \\
  \end{matrix} \right.
 \end{split}
\end{align}

where 

\begin{eqnarray} 
p_{th}=p^{(a.w.)}_{th} = \frac{1}{\left \lbrace 1 + (D-1)^{-(2-\frac{1}{N})} \right \rbrace} \nonumber \\   
\end{eqnarray} 

with $ D \geq 2 $ and $ N \geq 2 $. In $ 2 \otimes 2 $ case matrix (24) coincides with one from the example by Gisin and Peres (f-las(9-12) from Ref.[5]) for the case $ a=b=1/\sqrt{2} $ hence the obtained causal separability criterion of Eqs.(27,28) coincides in this case with the ppt-separability/entanglement criterion ( $0 \leq p \leq 1/2 $ -separable, $ 1 \geq p > 1/2 $- entangled) from Eq.(13) in Ref.[5]. Remarkably, in this "weakly mixing" N-free a-class  EC-situation the threshold value (28) of the entanglement parameter $ p $ has very weak dependence on the size of ensemble $ N $ (i.e. on the number of subsystems). Namely, the dependence $ p^{(a.w.)}_{th}(D) $ varies from $ 1/[1 + (D-1)^{-3/2}] $ in the "minimal" case $ N=2 $ (only two quantum subsystems of dimensionality $ D $ each) to $ 1/[1 + (D-1)^{-2}] $ in the limit $ N \rightarrow \infty $ (very large or infinite ensemble of equally connected $ D $-dimensional quantum systems weakly mixed and decoupled from each other at the moment of measurement of respective density matrix elements). Therefore, one can conclude that in the case of "weak mixing" the ensemble of $ N $-free equally connected quantum subsystems with no discrimination in mixing between different configurations of subsystems' base quantum states (the a-class of density matrices) \textit{becomes asymptotically separable in the limit $ D \rightarrow \infty $, i.e. in the case if all subsystems of given ensemble have infinite discrete or continuous spectra of observable eigenvalues}. The latter statement is also intuitively clear: any such ensemble is in general weakly coupled (even at the moment of a given entanglement preparation in system's past) due to pre-factors $ 1/(D-1)^{N} $ in (24), thus, in the limit $ D \gg 2 $  such the ensemble would have really vanishing degree of entanglement and therefore, all the subsystems of the ensemble would become asymptotically independent (or disentangled) from each other in the limit $ D \rightarrow \infty $. As it follows from (28) for large ensembles $ N \gg 2 $ this tendency of asymptotic statistical independence of constituent subsystems is a bit stronger than for small enough ensembles ( $ N \geq 2 $). This is because the existing small degree of entanglement is more "dissolved"  in the larger ensemble. As well the related entanglement/separability criterion of  Eqs.(27,28)  has a wide number of applications for a variety of different quantum systems consisting of equivalent quantum subsystems being equally connected to each other in their past but then decoupled by the moment of their measurement.

Analogously, substituting the respective matrix elements from Eq.(25) into Eq.(19) for a generic (non-interacting, i.e. N-free, see below) weakly mixing type of b-class of EC-density matrices  one obtains for the $ j $-th  configuration being $ m $ times pt-transformed (see details in the Subsection 4 of the Appendix A)

\begin{eqnarray} 
\nonumber
W^{(b.w.)}_{m(j);m}(p)=  (1 - p )^{2m(j)} p^{2(N-m(j))} \\ \nonumber
\times \left\lbrace  1 -  \frac{1}{(D-1)^{2N-1}} \left(\frac{1}{p} - 1 \right)^{-2m}  \right\rbrace. \nonumber \\
\nonumber \\
 m=\pm1,\pm2,..,\pm(N-1); \ m(j)=1,2,..,N.  \ \ \ \    
\end{eqnarray} 

Unlike the previous a-class in present b-class case from Eq.(29) it is evident that given EC-density matrix is \textit{absolutely separable} only in the case if $ W^{(b.w.)}_{m(j);m}(p) \geq 0 $ for all values of $ m $ and $ m(j) $. Otherwise, such density matrix will be entangled due to causal criterion (19-22). Especially, solving inequalities (22) with respect to expression (29) one obtains

\begin{align}
\begin{split}  
\left\{ 
 \begin{matrix}
  p_{th1}(\vert m \vert) \leq p  \leq p_{th2}(\vert m \vert) , \  \ \vert m \vert-separable    \\  
\\ 
  otherwise, \  \ \vert m \vert-entangled  \\
  \end{matrix} \right.
 \end{split}
\end{align}

where 

\begin{eqnarray} 
p_{th1}(\vert m \vert)=p^{(b.w.)}_{th1}(\vert m \vert) = \frac{1}{\left \lbrace 1 + (D-1)^{\frac{(2N-1)}{2\vert m \vert}} \right \rbrace} \nonumber \\   
\end{eqnarray} 

and

\begin{eqnarray} 
p_{th2}(\vert m \vert)=p^{(b.w.)}_{th2}(\vert m \vert) = \frac{1}{\left \lbrace 1 + (D-1)^{-\frac{(2N-1)}{2\vert m \vert}} \right \rbrace} \nonumber \\   
\end{eqnarray} 

with $ \vert m \vert=1,2,..,(N-1) $ , $ D \geq 2 $ and $ N \geq 2 $. One can see, that for weakly mixed b-class N-free EC-density matrices the causal separability criterion of Eqs.(30-32) is fulfilled within the narrow enough interval of $ p $ values which is confined between two thresholds of Eqs.(31,32), thus, forming the \textit{separability windows} for any number $ m(j) $ of pt-transformed subsystems involved into the $ j $-th configuration of the system. Similar "separability windows" in the parameters' space of multi-partite quantum system have been revealed in recent study on bound entanglement of isotropic quantum states \cite{35} (see Fig.2 in Ref.\cite{35}).

Thus, unlike the a-class case in the b-class situation the separability or entanglement of a given weakly mixed N-free EC-density matrix depends on the number $\vert m \vert $ of subsystems those having experienced the LCR-transformation. Therefore, the criterion (30-32) is different for the entanglement/separability of different (with respect to number $ \vert m \vert$ of subsystems involved) sub-ensembles with the rest of $ N $-partite ensemble of given b-class EC-density matrix. Remarkably, both thresholds of Eqs.(31,32) coincide in the situation $ D=2 $ where one has $ p^{(b.w.)}_{th1}=p^{(b.w.)}_{th2}=1/2$ and hence any weakly mixed N-free b-class ensemble of $ N $ equally connected  $ D=2 $-dimensional quantum subsystems (qubits, $ 1/2 $-spins, etc.) is separable only if $ p=1/2 $ and is entangled otherwise. This fact coincides with the criterion has been found earlier by Horodeckis for $ 2 \otimes 2 $ realization of this situation \cite{6}. 

From Eqs.(31,32) one sees that if $ \vert m \vert=\vert m \vert_{min}=1 $ then one has

\begin{eqnarray} 
p^{(b.w.)}_{th1}(1) = \frac{1}{\left \lbrace 1 + (D-1)^{(N - \frac{1}{2})} \right \rbrace} \nonumber \\   
\end{eqnarray} 

and

\begin{eqnarray} 
p^{(b.w.)}_{th2}(1) = \frac{1}{\left \lbrace 1 + (D-1)^{-(N - \frac{1}{2})} \right \rbrace} \nonumber \\   
\end{eqnarray} 

while the opposite limiting case when $ \vert m \vert = \vert m \vert_{max}=(N-1) $ yields

\begin{eqnarray} 
p^{(b.w.)}_{th1}(N-1) = \frac{1}{\left \lbrace 1 + (D-1)^{\left(1+\frac{1}{2(N-1)}\right)} \right \rbrace}  \nonumber \\   
\end{eqnarray} 

and

\begin{eqnarray} 
p^{(b.w.)}_{th2}(N-1) = \frac{1}{\left \lbrace 1 + (D-1)^{-\left (1+\frac{1}{2(N-1)}\right )} \right \rbrace}. \nonumber \\   
\end{eqnarray} 

From Eqs.(33,34) it follows that if one has very big (or infinite) ensemble of $ N \rightarrow \infty $ identical quantum subsystems  non-interacting with each other (i.e. N-free) at the stage of measurements performed on a given b-class EC-density matrix, where each subsystem has a finite number $ D > 2 $ of observable $ k $ eigenstates then threshold values (33,34) yield $ p^{(b.w.)}_{th1}(1)_{\vert N \rightarrow \infty } \rightarrow 0  $ and $ p^{(b.w.)}_{th2}(1)_{\vert N \rightarrow \infty } \rightarrow 1  $ -i.e.  one can identify each subsystem in corresponding ensemble as one being \textit{asymptotically separable or statistically independent} from the rest of the system at any value of the entanglement parameter $ p $. Interestingly, the latter statement is not symmetric for the degree of entanglement between  $ N-1 $ subsystems of this kind and the remaining subsystem from the same ensemble, where from Eqs.(35,36) one has  $ p^{(b.w.)}_{th1}(N-1)_{\vert N \rightarrow \infty}  \rightarrow \frac{1}{D}$ and $ p^{(b.w.)}_{th2}(N-1)_{\vert N \rightarrow \infty}  \rightarrow 1 - \frac{1}{D}$ for respective threshold values. Therefore, if given weakly mixing N-free b-class  EC-density matrix encodes any finite or infinite number $ N $ of subsystems ($ N \geq 2 $), where each constituent subsystem has very many (or infinite) number of its base states (i.e. in the limit $ D \rightarrow \infty $ )   then such subsystems are also \textit{asymptotically absolutely separable} at any value of entanglement parameter $ p $. On the other hand, from Eqs.(33-36) it follows that any number of qubits, e.g. $ 1/2 $-spins from any (large or small) ensemble, where all $ N $ its elements are connected with each other via weakly mixing N-free b-class EC-density matrix - is \textit{entangled} with the rest of the system at any value of entanglement parameter $ p $ except the thresholds value itself $p=p^{(b.w.)}_{th1}(D=2) = p^{(b.w.)}_{th2}(D=2) =1/2 $, where the system remains absolutely separable.
    
In fact, one can understand weakly mixing case as one where constituent quantum subsystems of the entire quantum system are strongly thermalized via their common environment (thermostat), or, alternatively, such quantum subsystems were interacting with each other in the past, at the stage of given density matrix preparation and such interaction had time to "die off" till the moment of interest. Anyway, as the result, in the latter case in each among two limiting situations: $ N \gg 2 $ and/or $ D \gg 2 $ the pre-factors $ \frac{1}{(D-1)^{N}} $ in all the off-diagonal matrix elements all tend to zero for large enough statistical ensembles ($ N \gg 2 $ - e.g. for "long" quantum spin chains ) and/or for large dimensionality of subsystems' Hilbert subspaces (for $ D \gg 2 $, e.g. for the systems similar to one of several (or infinitely many) weakly coupled quantum harmonic oscillators at temperatures near absolute zero). Therefore, weakly mixing case of large enough EC-density matrices describes a very small mixing between all completely orthogonal configurations of the system with respect to eigenvalues of certain local observable due to the weak off-diagonal coupling of the system to the external degrees of freedom  or due to the weak interaction between the parties of the ensemble (system) in its past. 

In contrary, in the opposite strongly mixing case, the renormalisation pre-factor  $ \frac{1}{(D-1)^{N}} $ is absent in all the off-diagonal matrix elements and, hence, the "mixing" between all completely orthogonal configurations of the system (ensemble) with respect to eigenvalues of certain local observable being measured - is strong and independent on the size $ N $ of the ensemble as well as on the number $ D $ of base states in each subsystem. Physically, the case of "strong mixing" in both a) and b)-classes of EC-density matrices models a strong off-diagonal coupling between all constituent subsystems through their common environment in the past and/or in present. Here the former (N-free) case obviously means strong interaction between subsystems at the stage of system preparation in the past whereas the latter (N-coupled) limit refers to strongly interacting regime of subsystems' collective quantum dynamics. This is because both these sorts of strong off-diagonal coupling are the sources of either strong quantum fluctuations (due to coupling to environment) or strong quantum correlations in the dynamics of different parties (e.g.due to specific ensemble preparation). And, in turn, strong quantum fluctuations and/or strong quantum correlations between the subsystems both can create a "strong mixing" of different completely orthogonal configurations within the quantum state of a whole quantum system in its $ D^{ \otimes N} $ Hilbert space.

As the result, quantity $ W^{(a.s.)}(p) $ for "strongly mixing" a-class N-free EC-case takes the form (for the details see Subsection 2 of the Appendix A)

\begin{eqnarray} 
W^{(a.s.)}(p)=  \vert p \vert^{N} \left\lbrace  (1 - \vert p \vert )^{N} -  (D-1) \vert p \vert^{N}  \right\rbrace. \nonumber \\   
\end{eqnarray}  

Hence, the separability/entanglement criterion preserves its form of inequality (27) though with entirely different threshold value $ p^{a.s.}_{th} $ as compared to Eq.(28) 

\begin{eqnarray} 
p_{th}=p^{(a.s.)}_{th} = \frac{1}{\left \lbrace 1 + (D-1)^{\frac{1}{N}} \right \rbrace} \nonumber \\   
\end{eqnarray} 

with $ D \geq 2 $ and $ N \geq 2 $. Substituting Eq.(38) into inequalities (27) one obtains desired causal separability/entanglement criterion for the strongly mixing N-free a-class of equally connected density matrices of arbitrary dimensionality. According to Eq.(38) in the simplest case $ D=2 $ and $N=2 $ the criterion (27,38) again reduces to the ppt-criterion for quantum state from Gisin and Peres example \cite{5} since in this case corresponding $ 4 \times 4 $ density matrix reduces to one described by Peres-Horodecki separability criterion \cite{5,6} . In general, equation (38) means that for any finite discrete $ D $ which satisfies $2 \leq D \ll N  $ the corresponding N-free EC-ensemble (characterized by its entanglement parameter $ p $ ) remains entangled in  a wide range of $ p $ values between $ 1/2 $ and $ 1 $ and separable for any $ p $ between $ 0 $ and $ 1/2 $) analogously to the situation of  $ 4 \times 4 $ ppt-case from Ref.[5]. Thus, one can call this situation as the \textit{asymptotical Werner-entangled} one since the causal separability criterion (CSC) for such type of quantum system is reduced to the ppt- criterion for two-qubit Werner-coupled system (see e.g. \cite{5,6,8} ). However, for large enough dimensionalities of involved subsystems if $ 2 \leq N \ll D  $ the separability of corresponding strongly mixed a-class N-free EC-ensembles holds only in the infinitesimally small vicinity of the value $ p=0 $ while basically the entire interval of possible values of parameter $ p $ corresponds to situation of entangled ensemble. The latter situation we will call as one being \textit{asymptotically entangled}. More precisely, one has for the "strongly mixing" variation of N-free a-class equally connected (EC-) case

\begin{align}
\begin{split}  
\left\{ 
 \begin{matrix}
  N \ll \ln(D-1) \  \ -asymptotically \ entangled \    \\  
\\ 
 \ln(D-1) \ll N \ -asymptotically \ Werner-entangled \   \\
  \end{matrix} \right.
 \end{split}
\end{align}

From Eqs.(27,38) and Eqs.(39) it follows that strongly mixing N-free a-class equally entangled ensemble of arbitrary number of qubits, say, for $ N $  $ 1/2 $-spins ($ D=2 $) remains Werner-entangled at any number $ N \geq 2 $. Whereas, the separability/entanglement regions of $ p $ values for the same kind of ensembles but for any spin eigenvalues being higher than $ 1/2 $ (i.e. for larger values of $ D $: $ D=3,4,.. $ ) -start to depend on the number $ N $ of subsystems in the ensemble more substantially. Especially, for large enough $ D \geq 3 $ the increase of $ N $ gradually "relaxes" the entanglement criterion (27,38) resulting in the asymptotically Werner-entangled situation in the limit $ N \rightarrow \infty $. Interestingly, the realisation of similar (asymptotically Werner-entangled) regime for uniformly strongly interacting quantum subsystems with $ N \gg D $ has been recently demonstrated in the recent experiment \cite{27} where the entanglement  between two ensembles of interacting atoms has been measured as the function of the parameter of inter-ensembles interaction  (see Fig.4 in Ref.\cite{27}).

The same effect of entanglement restriction "relaxation" with the increase of $ N $ holds also in the opposite case of large enough ensembles of the subsystems with $ D \gg 3 $. Whereas for any finite strongly mixed N-free equally connected ensembles with $ D \rightarrow \infty $ (such as e.g. two or more quantum oscillators being coupled and than - decoupled from each other in the past and as well, two or more quantum systems with continuous spectra of observable eigenvalues which were interacting in the past) the Eqs.(38,39) demonstrates the \textit{asymptotic entanglement} of such subsystems at any value  $ N \geq 2 $. Although, interestingly, the equally connected infinite ensemble of quantum oscillators or infinite ensemble of systems with continuous spectra of certain observable eigenvalues both \textit{remain asymptotically Werner-entangled} according to Eqs.(27,38) for such kind of equal entanglement in density matrices.

For the strongly mixing (N-free) b-class of EC-density matrices, in the same way, as it has been done for Eqs.(29-32), one can obtain the causal separability/entanglement criterion just by means of the renormalisation $ \frac{p^{2N}}{(D-1)^{2N}} \rightarrow p^{2N} $ in the Eq.(A11) from the Appendix A, while Eq.(A10) should remain as it is. As the result, analogously to Eq.(29) one obtains

\begin{eqnarray} 
\nonumber
W^{(b.s.)}_{m(j);m}(p)=  (1 - p )^{2m(j)} p^{2(N-m(j))} \\ \nonumber
\times \left\lbrace  1 -  (D-1) \left(\frac{1}{p} - 1 \right)^{-2m}  \right\rbrace. \nonumber \\
\nonumber \\
 m=\pm1,\pm2,..,\pm(N-1); \ m(j)=1,2,..,N.  \ \ \ \    
\end{eqnarray}

Therefore, for strongly mixing N-free equally connected density matrices of b-class (i.e. for ones with certain discrimination in quantum transitions between different configurations of subsystems' base states) one has the same separability/entanglement causal criterion of Eq.(30), though with different values of both thresholds. Especially, substituting Eq.(40) into condition (30) one yields

\begin{eqnarray} 
p_{th1}(\vert m \vert)=p^{(b.s.)}_{th1}(\vert m \vert) = \frac{1}{\left \lbrace 1 + (D-1)^{-\frac{1}{2\vert m \vert}} \right \rbrace} \nonumber \\   
\end{eqnarray} 

and

\begin{eqnarray} 
p_{th2}(\vert m \vert)=p^{(b.s.)}_{th2}(\vert m \vert) = \frac{1}{\left \lbrace 1 + (D-1)^{\frac{1}{2\vert m \vert}} \right \rbrace} \nonumber \\   
\end{eqnarray}

with $ \vert m \vert=1,2,..,(N-1) $ , $ D \geq 2 $ and $ N \geq 2 $. Substituting Eqs.(41,42) into criterion (30) we see that in the strongly mixing case any b-class N-free EC-density matrix encodes the ensemble of \textit{absolutely entangled, i.e. entangled at all possible values of $ \vert m \vert $} quantum subsystems of dimensionality $ D > 2 $ each, while at $ D=2 $ it is separable only at one special point where $ p=p^{(b.s.)}_{th1}(D=2)=p^{(b.s.)}_{th2}(D=2)=1/2 $ remaining entangled at all other possible values of entanglement parameter, as it was predicted previously for the example of b-class EC-density matrix in $ 2 \otimes 2 $ Hilbert space given by Horodeckis \cite{6}. As well, from Eqs.(30,41,42) it follows that  for \textit{any} $ N \geq 2 $ and $ D > 2 $ any strongly mixing b-class N-free equally connected density matrix \textit{is always entangled}. Therefore, among the ensembles of different quantum subsystems connected with each other by means of strongly mixing b-class N-free EC-density matrices only those consisting of $ 1/2 $-spins or qubits ($ D=2 $) are separable at single entanglement parameter value $ p=1/2 $,  while \textit{all ensembles of such type with $ D > 2 $ remain entangled}.

At last, in each of the above mentioned distinct cases all $ N $ subsystems which are described by means of respective EC-density matrix can be either \textit{decoupled from each other (though possibly being interacting at the preparation stage of density matrix evolution in the past)} or \textit{interacting with each  other at the moment of density matrix measurement}. These two possibilities one can characterize by means of two respective limits in the true dimensionality of system's Hilbert space and in the number of available virtual quantum transitions between different completely orthogonal configurations of the system. The "non-interacting" case here has the limit which I called as the "\textit{N-free}" case, where different subsystems of a given N-partite quantum system "do not talk to each other" at the moment of measurement (hence those are just elements of the  statistical ensemble) and, therefore, any virtual quantum transitions between base states of different subsystems are forbidden in this case. The maximal dimensionality of the available Hilbert space for such N-free quantum systems (ensembles) is $ D^{N} $, whereas, the number of completely orthogonal configurations for one configuration being chosen is $ (D-1)^{N} $, and the number of simultaneous virtual quantum transitions which may occur between two completely orthogonal configurations of the ensemble is the same as the number of such virtual quantum transitions for one constituent subsystem, i.e. it is equal simply to $ (D-1) $. 

On the other hand, in the opposite "interacting" case one has the situation where given density matrix describes N-partite quantum system of $ N $ constituent quantum subsystems interacting with each other at the moment of measurement of density matrix elements (via the measurement of eigenvalues of corresponding local observable in each subsystem). Here, in opposite to the former, one has the limit, which I called as the "\textit{N-coupled}" case. For such sort of interacting systems all types of virtual (and real as well) quantum transitions (say, tunnelling) between all parties become possible \cite{16,17,18,19,24,25} . Therefore, the true dimensionality of system's Hilbert space can be much lesser than $ D^{N} $ because the eigenstates of different subsystems can now overlap, and in the limit of uniform strong coupling between all subsystems such the dimensionality can reach its minimal value being equal to $ D $- the dimensionality of Hilbert subspace for just one constituent subsystem, while all the rest $ D^{N-1} $ base quantum states of the entire system (or ensemble) can be generated as the result of strong interaction (e.g. tunnelling) between all the constituent subsystems. Thus, in such the case the number of configurations of subsystems' eigenstates being completely orthogonal to a certain chosen configuration will be simply $ (D-1) $, whereas the number of possible virtual quantum transitions to all completely orthogonal (when decoupled) configurations in this case should be equivalent to the number of all variants of tunnelling between $ (D-1) $ base states of $ N $ subsystems, i.e. to the number $ (D-1)^{N} $.

Therefore (see the details in the Subsection 3 of Appendix A)

\begin{eqnarray} 
W^{(a.w.)}_{(I)}(p)= \frac{\vert p \vert^{2N}}{(D-1)^{N}} \left\lbrace (D-1)(1 - \vert p \vert )^{N} -  \vert p \vert^{N}  \right\rbrace. \nonumber \\   
\end{eqnarray}
 
 Hence, in the interacting (N-coupled case) of weakly mixing a-class the EC-ensemble one has the same criterion (27) of separability/entanglement but with different threshold value $ p_{th}=p^{a.w.}_{th(I)} $ of the entanglement parameter 

\begin{eqnarray} 
p_{th}=p^{(a.w.)}_{th(I)} = \frac{1}{\left \lbrace 1 + (D-1)^{-\frac{1}{N}} \right \rbrace} \nonumber \\   
\end{eqnarray} 

with $ D \geq 2 $ and $ N \geq 2 $. Obviously, formula (44) is \textit{dual} to Eq.(38) for the threshold value in the N-free (non-interacting) case of strongly mixing EC-density matrix (a broader context of such the duality will be discussed below). Especially, one can transform these two dual threshold values (38) and (44) into each other by means of a simple interchange   $ (D-1) \leftrightarrow (D-1)^{-1} $ of the argument $ (D-1) $ in both threshold values $ p^{(a.w.)}_{th(I)} $ and $ p^{(a.s.)}_{th} $. As the result, inequality of the type of Eq.(39) in the weakly mixing N-coupled (interacting) case will modify as compared with the strongly mixing N-free (non-interacting) case of Eq.(39). At $ N \ll \ln(D-1) $  the states which are encoded by N-coupled weakly mixing EC-density matrix become \textit{asymptotically separable} in contrary with the corresponded inequality in Eq.(39) for the opposite completely entangled state. Whereas, for the regime with $ N \gg \ln(D-1) $ the degree of entanglement between subsystems of N-coupled weakly mixing a-class EC-density matrix is the same as one for Werner-entangled two-qubit system similarly to the regime of the validity of the same inequality (39) in the dual N-free strongly mixing a-class case of EC-density matrix.

In order to obtain the entanglement/separability criterion for strongly mixed variation of arbitrary  N-coupled a-class EC-density matrices analogously to previous N-free case one needs only to make the replacement $ p^{2} \rightarrow (D-1)p^{2}$ in Eq.(A7) in the Subsection 3 of the Appnedix A while Eq.(A6) there remains the same: $ P^{(a.s.)}_{\varnothing(I)}(p)=P^{(a.w.)}_{\varnothing(I)}(p) $ and $ P^{(a.s.)}_{\circlearrowright(I)}(p) = (D-1)^{N}\vert p \vert^{2N}$. This yields

\begin{eqnarray} 
W^{(a.s.)}_{(I)}(p)=  \vert p \vert^{N} \left\lbrace  \frac{(1 - \vert p \vert )^{N}}{(D-1)^{N-1}} -  (D-1)^{N}\vert p \vert^{N}  \right\rbrace. \nonumber \\   
\end{eqnarray} 

Equation (45) results in the separability/entanglement criterion in the form of Eq.(27) with following threshold value  $ p_{th}=p^{a.s.}_{th(I)} $ of entanglement parameter

\begin{eqnarray} 
p_{th}=p^{(a.s.)}_{th(I)} = \frac{1}{\left \lbrace 1 + (D-1)^{2-\frac{1}{N}} \right \rbrace} \nonumber \\   
\end{eqnarray} 

with $ D \geq 2 $ and $ N \geq 2 $. Obviously, separability threshold (46) for strongly mixed case is \textit{dual} to separability threshold (28) for a-class strongly mixing N-free EC-density matrices. The threshold (46) only weakly differs from one for weakly mixed case for the same a-class of N-coupled EC-density matrices since presumed interaction affects the system similarly  to "strongly mixing" off-diagonal matrix elements of a-class EC-density matrix. 

Thresholds (44,46) for a-class equally connected interacting density matrices have one interesting peculiarity. In the limit $ N \rightarrow \infty $ for any finite $ D $ the ratio $ N_{o.c.}/N_{v.t.}=D/D^{N}=D^{-(N-1)} $ tends to zero since in the strongly interacting N-coupled case the true number of many-body eigenstates of the ensemble will be equal to $ D $ by assumption, i.e. the latter is finite unlike the number $ N $ of involved subsystems. According to the standard many-body localization (MBL-) theory this corresponds to the situation where our interacting quantum system (or interacting ensemble) is \textit{non-ergodic} being "localized" as a whole within a finite number of its base states in the infinite-dimensional Hilbert space of the entire infinite ensemble \cite{17} . In other words, the interacting case of interest for big enough or infinite quantum ensembles deals with so-called \textit{many-body localized (or MBL-)} quantum states encoded in chosen EC- type of density matrices in the "fully interacting" or N-coupled case. On the other hand, from the threshold values (44,46) of entanglement parameter in the N-coupled case of interest one can see that in the limit of infinite interacting ensembles $ N \rightarrow \infty $ these two thresholds both become asymptotically independent on $ N $ being equal to $ 1/2 $ for "weakly mixing" case and to $ [1 + (D-1)^{2} ]^{-1}  $ for "strongly mixing" situation (with $ D \geq 2 $ everywhere). Therefore, one may conclude from the latter that \textit{interacting quantum many-body system with MBL-quantum states has a degree of entanglement being equal to degree of entanglement of its MBL-base states}. Remarkably, for the limit where both $ D $ and $ N $ tend to infinity (i.e. for infinite ensembles of subsystems with infinite or continuous spectra) , while the inequality $ N \gg D $ is kept, one can claim that \textit{any infinite strongly mixed MBL-quantum ensemble of subsystems with infinite discrete or continuous spectra of subsystems' eigenvalues is entangled, while in the "weakly mixed" case such infinite ensemble has the same degree of entanglement as any two-qubit Werner-entangled system has at the same value of its entanglement parameter $ p $ }.
  
As for the interacting (N-coupled) EC-density matrices of b-class, here, analogously to previous (N-free) b-class of Eqs.(40-42) one can obtain (see details in the Subsection 5 of the Appendix A)

\begin{eqnarray} 
\nonumber
W^{(b.w.)(I)}_{m(j);m}(p)=  \frac{(1 - p )^{2m(j)} p^{2(N-m(j))}}{(D-1)^{N-1}} \\ \nonumber
\times \left\lbrace  1 -  \frac{1}{(D-1)} \left(\frac{1}{p} - 1 \right)^{-2m}  \right\rbrace. \nonumber \\
\nonumber \\
 m=\pm1,\pm2,..,\pm(N-1); \ m(j)=1,2,..,N.  \ \ \ \    
\end{eqnarray} 

From the comparison of Eqs.(40) and (47) one can easily conclude that separability/entanglement criterion for weakly mixed b-class EC-density matrices being N-coupled preserves its form (30) for N-free EC-density matrices of such type with simple exchange of two threshold values of Eqs.(41,42). Namely, for the particular case of interest one has $ p^{(b.w.)}_{th1(I)}(\vert m \vert)=p^{(b.s.)}_{th2}(\vert m \vert) $ and $ p^{(b.w.)}_{th2(I)}(\vert m \vert)=p^{(b.s.)}_{th1}(\vert m \vert) $, correspondingly. Thus, one can see that in the interacting N-coupled case in the limit $ m,N \rightarrow \infty $ and $ D > 2 $ the system tends  to remain separable in the entire region of the entanglement parameter values. Whereas, for any finite $ m,N $ for example for $ m=1 $ at \textit{any} $ N $ having $ D > 2 $ the "separability window" between the values  $ \frac{1}{\left \lbrace 1 + (D-1)^{\frac{1}{2}} \right \rbrace} $ and $  \frac{1}{\left \lbrace 1 + (D-1)^{\frac{-1}{2}} \right \rbrace} $ of the entanglement parameter still exists in the system, while for the rest of entanglement parameter values such kind of quantum system remains entangled. At the same time, in the limit $ D \rightarrow \infty $ (i.e. for  weakly mixed b-class equally connected ensemble of any number of quantum oscillators (for $ N \ll D$ though) the latter subsystems (quantum oscillators) tend all to be \textit{asymptotically separable} at any value of the entanglement parameter $ 0 \leq p \leq 1 $.

Finally, for the strongly mixed N-coupled b-class of EC-density matrices, corresponding causal separability/entanglement criterion for the entanglement parameter $ p $  can be obtained by previously used renormalisation $ \frac{p^{2N}}{(D-1)^{2N}} \rightarrow p^{2N} $ in Eq.(A13) leaving Eq.(A12) intact. Then the result for $  W^{(b.s.)(I)}_{m(j);m}(p)$ reads

\begin{eqnarray} 
\nonumber
W^{(b.s.)(I)}_{m(j);m}(p)=  \frac{(1 - p )^{2m(j)} p^{2(N-m(j))}}{(D-1)^{N-1}} \\ \nonumber
\times \left\lbrace  1 - (D-1)^{2N-1} \left(\frac{1}{p} - 1 \right)^{-2m}  \right\rbrace. \nonumber \\
\nonumber \\
 m=\pm1,\pm2,..,\pm(N-1); \ m(j)=1,2,..,N.  \ \ \ \    
\end{eqnarray} 

Surprisingly, formula (48) turns out to be \textit{dual} to Eq.(29) for $ W^{(b.w.)}_{m(j);m}(p) $ in N-free weakly mixing realization of b-class EC-density matrix. This is because Eq.(48) reduces to Eq. (29) by the replacement $ (D-1)^{(2N-1)} \rightarrow (D-1)^{-(2N-1)} $ in the second line of Eq.(48). As the result, separability/entanglement criterion for the N-coupled strongly mixing b-class of EC-density matrices has the same form (30) as one for N-free weakly mixing EC-density matrices of b-class, but with threshold values of Eqs.(31,32) being switched. Thus, for the former case one has for the entanglement parameter $ p $ the separability/entanglement criterion (30) with two threshold values $ p^{(b.s.)}_{th1(I)}(\vert m \vert)=p^{(b.w.)}_{th2}(\vert m \vert) $ and $ p^{(b.s.)}_{th2(I)}(\vert m \vert)=p^{(b.w.)}_{th1}(\vert m \vert) $. "Mapping" this replacement onto the explanations after Eqs.(31-36) in the above text one can arrive the remarkable conclusion that \textit{any strongly mixed N-coupled b-class equally connected density matrix always remains entangled except the case $ D=2 $(the two-qubits case of Ref.[6]) where such sort of EC-density matrix is entangled everywhere except one point where $ p=1/2 $ }.

In the above we have explored in details the application of the new causal separability/entanglement criterion of Eqs.(19-22) to the important wide class of one-parametric equally connected (EC-) density matrices. It is quite obvious that eight distinct situations  have been considered exhaust all the diversity of most important one-parametric (or equally Werner- connected) density matrices defined in $ D^{ \otimes N} $ Hilbert space. These eight cases include: two classes a) and b) corresponded to either equivalent or different off-diagonal matrix elements, where each class can be either interacting (N-coupled) or non-interacting (N-free) with two distinct sub-cases for each situation: weakly- or strongly mixing (i.e. corresponding to either small or big off-diagonal matrix elements as compared to diagonal ones). All these particular cases reveal those different  entanglement conditions which have their limiting cases for $ 4 \times 4 $ density matrices in the form of well-known ppt-separability criteria \cite{5,6}. This proves the universality of the novel causal separability/entanglement criterion being summarized in Eqs.(8,9) and Eqs.(17-22) for arbitrary density matrices.

\subsection{Dualities in the separability/entanglement thresholds for EC-density matrices}

At this point, one can see a remarkable duality between N-free and N-coupled regimes: to transfer in causal criterion (19) from one regime to another one needs just to switch the number of completely orthogonal configurations and the number of virtual quantum transitions (see also Eqs.(20,21)). Naturally, this peculiarity implies also the duality between the N-free (non-interacting) and N-coupled (interacting) variations of causal separability criterion of Eqs.(19-22). To see this for a)-class EC-density matrices one can compare the expressions (28) and (46) for the threshold values of entanglement parameter in the weakly mixing N-free and strongly mixing N-coupled cases, correspondingly. One can see that these two formulas can be reduced to each other by means of a simple transformation $ (D-1) \rightarrow (D-1)^{-1} $. The same remains true also for the threshold values of Eqs.(38) and (44) for strongly mixing N-free case and weakly mixing N-coupled cases. So, one can write down following general duality relation for function $ p_{th}( D-1 ) $ in all these cases

\begin{eqnarray} 
p^{(a.w.(s.))}_{th}\left( (D-1)\right)=p^{(a.s.(w.))}_{th(I)}\left((D-1)^{-1}\right). 
\end{eqnarray}

At the same time, for the b-class of EC-density matrices where one has two thresholds of separability  the consequences of the duality between N-free (non-interacting) and N-coupled (interacting) cases are even more pronounced. As it has been already mentioned in the previous subsection dual cases can be obtained just by switching values of two thresholds, i.e. $ p_{th1}(\vert m \vert) \Leftrightarrow p_{th2}(\vert m \vert) $. Therefore, such the duality for the functions $ p_{th1,2}(\vert m \vert) $ reads

\begin{eqnarray} 
p^{(b.w.(s.))}_{th1,2}(\vert m \vert) = p^{(b.s.(w.))}_{th2,1(I)}(\vert m \vert). 
\end{eqnarray}

\subsection{Application of the causal separability criterion for EC-density matrices: arrays of qubits}

First of all, here one can see one remarkable universality (or self-duality) of separability thresholds at $ D=2 $ (systems of equally entangled qubits or 1/2-spins): \textit{for arbitrary number $ N $ of uniformly connected qubits (i.e. in all situations where $ D=2 $ for both a- and b-classes, either interacting or not, either weakly or strongly mixing ensembles) the entanglement properties of the entire system remain the same as ones for respective sorts of equally connected density matrices acting in $ 2 \otimes 2 $ Hilbert space } \cite{5,6}. This is because there  exist only \textit{one} available completely orthogonal configuration for each configuration being chosen for the ensemble of arbitrary number $ N $ of quantum subsystems with two eigenstates each. Notice the difference with the predictions of Refs.\cite{11,12,21} where threshold values of the entanglement parameter for the ensembles of $ N $ qubits ($ D=2 $) are slight functions of $ N $, thus, providing better entanglement distillation in larger ensembles (i.e. at large $ N $) and, otherwise, a more wide separability region for smaller ensembles \cite{11,12,21}. The latter fact is  due to another type of non-universal separability criterion based on bi-partitions of density matrix  \cite{3}  being used  in Refs.[11,12] and, as well, due to the construction of underlying one-parametric density matrices in Refs.\cite{11,12} , using the basis of GHZ entangled quantum states of the ensemble (being mixed with white noise) \cite{26}  instead of the recurrent relations of the type of Eqs.(A1,A2) and Eqs.(A8,A9) (see Appendix A). As the result, it is obvious, that the separability criterion used e.g. in Refs.[11,12] for one-parametric density matrices with $ D=2 $ and $ N > 2 $ in the basis of the GHZ quantum states \cite{26}  involves not only those off-diagonal density matrix elements describing quantum transitions between completely orthogonal configurations of the system but also the off-diagonal elements which describes quantum transitions between some distinct configurations of the system where quantum state(s) of certain subsystem(s) do(es) not change during such the transition. Hence, one cannot assign any universal causal meaning to bi-partition entanglement criterion used in Refs.\cite{11,12} for qubit ensembles with $ D=2 $ and $ N > 2 $, thus, the corresponded predictions can be far enough from ones obtained by means of universal causal separability criterion (19-22) in this ($ D=2 $) case. However, it is quite remarkable that for quantum ensembles with $ D>2 $ and $ N \gg 2 $ (say, for array of qutrits) the causal separability criterion for the b-class of EC- density matrices gives the separability/entanglement regions of the entanglement parameter (see Eqs.(30-34) in the above) which sufficiently overlap with corresponded regions and separability conditions which were obtained in Refs.\cite{11,12,21} for similar type of physical systems. This can be due to the fact that given $ D=2 $ and $ N \gg 2 $ for the separability criteria used in \cite{11,12} for respective one-parametric density matrices (arrays of qubits) in the basis of GHZ quantum states one can "mimic" to some extent the consequences of the universal causal separability criterion (19-22) for physical systems with $ D > 2 $ and $ N > 2 $ (see e.g. Fig.1 and related comments in Ref.[11] ) . This is because at large enough $ N $ the relative contribution to the separability criterion of Refs. \cite{11,12} from all quantum transition amplitudes between non-completely orthogonal configurations of the system can be neglected as compared to ones between completely orthogonal configurations of the system, thus resulting in such the separability/entanglement criteria in the ensemble which turn out to be more similar to ones of the causal nature obtained from the universal causal separability criterion under consideration.

\subsection{Application of the causal separability criterion for EC-density matrices: ensembles of quantum oscillators}

As the result of the dualities (49,50), in the limit $ D \rightarrow \infty $ in both a)- and b)-class cases  in all region of the parameter $ p $ values respective \textit{weakly mixing N-free EC-density matrices} are \textit{separable}, whereas the \textit{strongly mixing N-coupled EC-density matrices} will be \textit{entangled} in the same region.  

It is quite easy to see that the latter prediction of a new causal separability criterion (19-22) for arbitrary EC-density matrices is also expectable physically. For example, it is obvious that any number of decoupled quantum harmonic oscillators or quantum many-body systems with continuous spectra of energy eigenvalues such as e.g. systems of decoupled quantum wires, etc. ( weakly mixing case at $ D \rightarrow \infty $) -represents separable quantum object since creation/annihilation operators of different quantum oscillators isolated from each other do commute \cite{14} . In contrary, we know that any detectable interaction between different quantum oscillators results in the hybridization of their eigenstates \cite{13,15}, as well even small coupling between two quantum wires causes the phenomenon of quantum tunnelling between the leads of such tunnel contact due to small but non-zero overlap of eigenstates corresponding to each lead \cite{18,20}. This is because two latter cases represent two examples of composite quantum systems, where related mathematical description maps on one in terms of the strongly mixing N-coupled equally connected density matrices. Thus, from the inequalities (27,30) together with duality relations of Eqs.(49,50) one can conclude that novel universal causal approach of Eqs.(16-23) to the density  matrices separability/entanglement criterion - \textit{highlights the phenomenon of quantum tunnelling in many-body interacting quantum systems from the perspective of the structure of underlying equally connected density matrices and configurations of base quantum states of the constituent quantum subsystems}.

In other two dual cases of a)-class EC-density matrices: N-free strongly mixing and N-coupled weakly mixing corresponding separability/entanglement criteria of inequalities (39) predict even more interesting situations with respect to the entanglement for some typical composite quantum systems. Both these cases (being dual to each other) in fact describe ensembles of strongly correlated quantum objects which have the same degree of entanglement between their different subsystems. In the N-free strongly mixing case such objects all have had considerable interaction with each other through the environment during all the time or only at the stage of system preparation in the past. Whereas, in the N-coupled though weakly mixing case such interaction between subsystems is much smaller tending to zero at $ D \rightarrow \infty $, however, since this case is called N-coupled the rank of the entire density matrix should to be equal to $ D $ thus, strong correlation between subsystems is presumed to take place in the past at the stage of preparation of a given density matrix or during the reduction of system's wave function at the stage of measurement of a given density matrix. The latter one can observe for example in any type of singlet quantum objects, e.g. being GHZ-correlated with each other \cite{26} . Obviously, any spin 1/2-singlet or a pair of EPR-correlated photons \cite{4}  could provide a "minimal" example of weakly mixing N-coupled a)-class equally connected density matrix with $ D=2 $ and $ N=2 $ (for more potential examples one can see e.g. \cite{1,2,4}) . On the other hand, one can easily imagine less trivial examples of strongly correlated quantum systems of this type: for example, a finite or infinite ensemble of quantum subsystems with finite spectrum each ($ D > 2 $, $ N \gg 2 $ or $ N \rightarrow \infty $), or a pair of remote quantum oscillators being prepared (in the past) in a certain superposition of their GHZ-correlated quantum states ($ N = 2$ and $ D \rightarrow \infty $ )\cite{13,15} . Then the separability or entanglement of such the systems can be described by means of inequalities (39). Remarkably, from the inequality dual to the first one among two of (39) one can conclude that it is impossible to create a GHZ-like quantum state of two quantum oscillators ( $ N=2 $, $ D \rightarrow \infty $) having vanishing interaction between these oscillators at the moment of measurement, even if they were prepared in a common correlated (or coherent) quantum state in the past. Hence, the quantum state of such a pair of equally connected quantum oscillators \textit{always will be separable}( a)-class weakly mixing N-coupled case of EC-density matrix). At the same time, non-vanishing interaction between these two quantum oscillators \textit{always results in the entangled state} according to the first inequality of (39) \cite{13,15} . The same is true also for any pair of quantum systems with infinite spectra of eigenvalues depending on the presence or absence of interaction between them, the examples are e.g. pairs of decoupled (or, in contrary, tunnel coupled) quantum wires, etc. (see e.g. \cite{16,17,18,20}).

\subsection{Application of the causal separability criterion for EC-density matrices: many-body interacting quantum systems and Anderson localization}

 Another interesting limiting case is the case of infinite number of quantum oscillators ($ D \rightarrow \infty $, $ N \rightarrow \infty  $ ) which usually models any quantum field at fixed space-time point. In this case  $ N \rightarrow \infty  $ is a number of different Fourier components of given quantum field  (i.e. a number of different secondary quantized creation/annihilation operators of quantum field eigenmodes), whereas $ D \rightarrow \infty $ represents infinite number of eigenstates for each quantized Fourier mode of the field. Here different quantum oscillators will represent different Fourier modes of quantum field in certain fixed space-time point. Then given $ N \ll D $ (at $ N, D \rightarrow \infty  $) from the inequalities (39) and dual to them, one obtains that in the absence of the off-diagonal (i.e. interaction-) terms in corresponding quantum field Hamiltonian one has infinitely many statistically independent Fourier modes (separable oscillators), whereas in the presence of coupling between different Fourier components of quantum field in the Hamiltonian one in general case obtains entangled quantum dynamics of respective quantum field. 
 
If instead one has a realization of uniformly coupled (or equally EC-) quantum ensemble with $ 0< p\ll \frac{1}{2}$ and $ N \gg 2 $ , $ D \gg 2 $, with $ N $ attributed to different spatial centers of particle scattering (say, the number of impurities in one-dimensional quantum wire) and $ D  $ being the number of distinct scattering states of the particle on each center (or site) then inequalities (39) bridge the famous Anderson localisation problem (see e.g. Refs.\cite{16,17} and references wherein). Namely in this situation entanglement between quantum states of a particle in the vicinity of different scattering centers (sites) means \textit{tendency of particle delocalisation} throughout the system, while a separability of such scattering states corresponded to different scattering centers - manifests \textit{the tendency of particle localization} in the vicinity of each scattering center (or impurity). Moreover, from the estimation of a crossover between asymptotically entangled and separable quantum states at $ 0 \leq p \ll \frac{1}{2} $ which are described by inequalities (39) one can obtain a remarkable crossover condition 

\begin{eqnarray} 
\frac{\ln(D-1)}{N_{cr}}  \sim 1, 
\end{eqnarray}

which separates asymptotically entangled situation at $ \ln(D-1) \gg N  $ for N-free strongly mixing a)-case from asymptotically Werner-entangled state of the same kind of systems at $ \ln(D-1) \ll N  $. Therefore, in order to transfer from asymptotically entangled to asymptotically Werner-entangled situation, where at $ 0< p\ll \frac{1}{2}$ the system becomes separable, one needs to increase $ N $ gradually having $ D $ fixed. Then crossover between separable and entangled regimes for arbitrary $ D \gg 2  $ will happen at the condition of Eq.(51). In Eq.(51) one could recognize a quite famous formula for the \textit{Anderson mobility edge} (for details one can see e.g. review paper of Ref.[17] ) if to interpret $ (D-1)=N_{ch} $ as the number of scattering channels per site and $ N= N_{s} $ as the number of excitations corresponded to distinct sites of one-dimensional Anderson model. In Ref.[17]  the numbers $ N_{ch} $ and $ N_{s} $ were defined as follows $ N_{ch}=W/V $ and $ N_{s}=W/(Vd) $ where $ W \rightarrow \delta_{\zeta}$ is typical energy mismatch for each virtual transition, $ V \rightarrow \lambda\delta_{\zeta}$ is nearest neighbour coupling matrix element and $ d \rightarrow T/\delta_{\zeta} $ is "coordination number" - a number of three-particle excitations to which given single-particle excitation is coupled, where $ T $ is the temperature. All this results in following mobility edge for Anderson metal-insulator transition \cite{17}
 
\begin{eqnarray} 
\frac{\lambda T_{cr}}{\delta_{\zeta}}\ln \left(\frac{1}{\lambda} \right)  \sim 1, 
\end{eqnarray}

where for all temperatures $ T > T_{cr}$ with $ T_{cr} $ from Eq.(52) the corresponding Anderson model is in the metallic (or delocalized) phase , while decreasing temperature below the critical value of Eq.(52) at $ T < T_{cr} $ one obtains a completely insulating (Anderson localized-) state of the entire system. Comparing Anderson mobility edge of Eq.(52) with the crossover condition of Eq.(52) one can see that gradual increase of $ N $ above its critical value from Eq.(51) at $ D \gg 2 $ results in the gradual transfer from the asymptotically entangled- strongly mixing N-free case to asymptotically separable weakly mixing N-coupled quantum state in the ensembles described by means of a)-class equally connected density matrices if to keep the corresponding entanglement parameter $ 0< p \ll \frac{1}{2} $. Hence, from Eqs.(51,52) it follows that Anderson metal-insulator transition \cite{17} with temperature decrease can be interpreted as the transition with the increase of  $ N $ from the asymptotically entangled to asymptotically separable quantum state of related Anderson model being encoded in the respective a)-class EC-density matrix from $ D^{\otimes N}  $ Hilbert space. 

\subsection{General predictions of causal separability criterion for arbitrary composite quantum systems}
 
As for the role of the size $ N $ of the ensemble in other cases of equally connected density matrices of both a)- and b)-classes, those are more trivial. For weakly mixing N-free cases of a)-and b) classes of EC-density matrices  the increase of $ N $  up to infinity can only "soften" separability/entanglement thresholds to the small degree. Thus, for weakly mixing N-free b-class of EC-density matrices this leads to the spread of the available "separability window" to its maximal possible size at $ N \rightarrow \infty $. (for examples of similar entanglement behaviour in multi-partite quantum systems one can see e.g. Refs.\cite{11,12,14}). Whereas the increase of parameter $ D \gg 2 $ for such EC-density matrices has stronger effect of the same kind on the separability thresholds and on the separability windows in both cases of a)- and b-classes weakly mixing EC-density matrices. 

Interestingly, for both strongly mixing N-free and dual weakly mixing N-coupled sub-cases of b-class EC-density matrices the ensemble size $ N $ does not affect directly the criteria of multi-partite separability or entanglement according to Eqs.(41,42) especially if one considers the separability conditions for just one party of the respective ensemble. The only effect of the separability windows growth here goes from the growth of the parameter $ D \geq 2 $ . Obviously, the obtained results at $ D \gg 2 $ correlate with entanglement/separability regions for isotropic quantum states being derived in Refs.\cite{28,29}.    

In general, from all the above, one can conclude that the requirements for the b)-class EC-density matrices to encode a certain separable quantum state turn out to be much stronger than ones for any EC-density matrices of a)-class. For example, in both (N-free and N-coupled) sub-cases of strongly mixing types of corresponded b)-class EC-density matrices  \textit{always remains entangled for any $ D > 2 $}, while for the same sub-cases of weakly mixing b-class EC-density matrices there exist certain "separability windows" for the values of entanglement parameter $ p $ at which the entire quantum system represents separable ensemble. These distinct entanglement properties of b)-class EC-density matrices are obviously connected with the fact of stronger "discrimination" between different completely orthogonal configurations in such systems as compared to the "non-discriminated" situation of the EC-density matrices of a)-class. However, a more clear physical background behind the entanglement properties of b)-class EC-density matrices as compared with ones of a)-class needs further investigation which is beyond our consideration in this research.  

Of course, in reality, arbitrary density matrix in $ D^{ \otimes N} $ Hilbert space does not necessary describe equally connected composite quantum system (or ensemble) which is governed by a single entanglement- (or coupling) parameter $ p $. Rather in the most general case one can speak about the most general form of causal  separability criterion (19-22) which is  defined by means of $ D^{2N} $ different (complex in  general) matrix elements of a given $ D^{N} $-dimensional density matrix with quantities $ K^{(D)}_{N} $ and $ \bar{K}^{(D)}_{N} $ from  Eqs.(14,15). Therefore, in the most general case the new universal causal separability/entanglement criterion of Eqs.(8,9) and Eqs.(17-22) can be difficult to calculate for $ N \gg 2 $ and/or $ D \gg 2 $ due to very large numbers of distinct- and completely orthogonal configurations (14,15) of subsystems' orthogonal base quantum states in the ensemble. For example, in the case where $ N \gg 2 $ and/or $ D \gg 2 $, in order to establish the fact of separability and/or entanglement of any constituent quantum subsystem within the larger quantum system or in the statistical ensemble one needs to analyse $ K^{(D)}_{N} \gg 1 $ equations of the form (19). This can represent a challenging computational task, since according to formula (14)  $ K^{(D)}_{N} $ increases exponentially with the growth of $ N $ - the number of constituent subsystems in given statistical ensemble. Nevertheless, all such complications one can overcome much more easily as compared to many existing calculation schemes\cite{3}, where one needs to calculate  the eigenvalues of pt-transformed matrices explicitly.  

Nevertheless, as it has been shown in the above on different examples of one-parametric equally connected density matrices, one can extract a lot of important information about the entanglement or separability for a wide range of different real quantum physical systems using all the above results on the entanglement/separability conditions for the one-parametric family of equally connected density matrices. This is because the majority of such composite quantum systems represent ensembles of constituent interacting identical subsystems being coupled with each other in a uniform fashion (such as e.g. ensembles of entangled qubits of different nature and most of strongly correlated interacting many-body quantum systems: cold atoms, tunnel contacts, spin chains, coupled quantum harmonic oscillators in nano-electromechanical systems, photon cavities, etc.)\cite{1,2,30,34}. 

Moreover, one can spread the applicability of all results obtained in the above for EC-density matrices even further, thus, covering a much more wide class of composite quantum systems, by means of the following considerations. Imagine one has arbitrary density matrix in its completely orthogonal configuration representation of Eq.(16) or Eq.(23). Then in the most common (i.e. \textit{not equally connected}) case one has  $ D^{N} $ different real numbers for respective diagonal matrix elements $ \rho_{\lbrace k \rbrace_{j} \vert \lbrace k \rbrace_{j}} $ in this representation, instead of just two real numbers $ p $ and $ (1-p) $ (and one entanglement parameter $ p $ with $ 0 \leq p \leq 1 $) in the former equally connected case. As well, in the most general situation calculation of causal separability criterion can involve $ 2 K^{(D)}_{N} \bar{K}^{(D)}_{N} $ distinct off-diagonal matrix elements $ \rho_{\lbrace k \rbrace_{j} \vert \lbrace \bar{k} \rbrace_{j'}} $ in Eq.(23) (the latter, generally speaking, all can be complex numbers) for all available quantum transitions between completely orthogonal configurations of the entire quantum system. Although for any realistic big enough density matrix (16) one needs to perform numerically all calculations of Eqs.(17-19) in order to check general causal condition (22), it is still possible to make some predictions on the character of the results one might obtain straightforwardly just approximating the most general situation by means of specific entanglement/separability conditions obtained for one-parametric equally connected type of density matrices. The goal here is to choose such the parametrization which would approximate real "non-EC" density matrix and, at the same time, which would result in a certain simple renormalization of the entanglement/separability thresholds obtained in the above for the family of EC-density matrices.

In fact, to make things easier, instead of the ensemble of $ N $ \textit{equal} quantum subsystems with \textit{unequal} probabilities of different base quantum states and unequal transition amplitudes between these base states in each subsystem, one can consider the ensemble of $ N $ \textit{unequally coupled} quantum subsystems (in the sense of matrix elements of the entire density matrix of the ensemble) where each subsystem when isolated from its surrounding would be described by $ D $-dimensional EC-density matrix of a)-class. Further, one can introduce the "hierarchy" of $ N $ different real numbers $ p_{k} $, $ k=1,2,..,N $ standing for diagonal matrix elements $ \rho_{\lbrace k \rbrace_{j} \vert \lbrace k  \rbrace_{j}} $ of density matrix in the configuration representation (16,23) as follows $ 0 \leq p_{N} <  .. < p_{k} < .. < p_{1} \leq 1   $ then consider quite reasonable case where for all off-diagonal elements between different configurations one has  $ \rho_{\lbrace k \rbrace_{j} \vert \lbrace \bar{k} \rbrace_{j'}}=\gamma (\lbrace k \rbrace_{j}) \rho_{\lbrace k \rbrace_{j} \vert \lbrace k \rbrace_{j}} $ with $ \gamma (\lbrace k \rbrace_{j})$ being  analogous to the quantum transition amplitudes which depend on the specific configuration $  \lbrace k \rbrace_{j} $ of all subsystems' base states. Let us choose the parametrization $ \gamma (\lbrace k \rbrace_{j})= \gamma = const \ll 1 $ and $ \rho_{\lbrace k \rbrace_{j} \vert \lbrace k  \rbrace_{j}} = p_{k} = p/k $, $ k=1,2,..,N $. Then from the basic equations (16-23) it follows that such the case is reduced to the a)-class of EC-density matrices though  with the renormalized entanglement thresholds $ p_{th(r)} $ as compared to corresponding thresholds $ p_{th(ee)} $ in the a)-class EC- situation

\begin{eqnarray} 
\nonumber
 p_{th(EC)} = \frac{1}{1+ (D-1)^{\alpha(N)}} \\
 \Longrightarrow p_{th(r)} = \frac{1}{1+ \left( \frac{\gamma m!}{N!} \right)^{\frac{1}{N}}(D-1)^{\alpha(N)}},
\end{eqnarray}

with $ m=1,2,..,N $ dependent on the concrete system's configuration and $ \alpha(N) $ being the ensemble size (i.e. $ N $ -)-dependent function which can be approximated as corresponded power from different sub-cases of a)-class equally entangled density matrices. Thus, the model resulting in thresholds renormalization of Eq.(53) for a)-class equally connected density matrices evidently represents the simplest approach to the unequally connected or just arbitrary density matrices. From Eq.(53) one can see the most robust effects of unequal connection or coupling between different subsystems. 

First of all, from Eq.(53) it is evident that the renormalization function  $ f(m,N)=\left( \frac{\gamma m!}{N!} \right)^{\frac{1}{N}} $ is confined between $ \left( \frac{\gamma}{N!} \right)^{\frac{1}{N}} \simeq  \frac{\gamma^{\frac{1}{N}}}{N}$ for $ m=1 $ and $ \left( \frac{\gamma}{N} \right)^{\frac{1}{N}} $ for $ m=N-1 $. Therefore, in the limit $ N \rightarrow \infty $ the renormalization function $ f(N-1,N) $ tends to one - meaning \textit{no renormalization} for $ m=N-1 $, as compared to the corresponding a)-class of equally connected case, whereas for $ m=1 $ in the same limit one has $ f(1,N \rightarrow \infty) \rightarrow 0 $ meaning the \textit{considerable increase of the entire separability region, as compared to respective a)-class of equally connected density matrices}. The latter observation shows that \textit{in general, the unequal connection between constituent subsystems of the ensemble enlarges separability of any constituent subsystem of the ensemble, while the separability properties of the rest $ N-1 $ subsystems remains insensitive to the difference in connections between the subsystems}. 

However, in the case of the arbitrary $ D $-dimensional blocks joined into single arbitrary $ D^{N} $-dimensional density matrix one cannot extract any simple enough analytical expressions for the estimations of corresponding separability/entanglement thresholds, hence, in the most general case on should use also the most general form of the causal separability criterion of Eqs.(19,22). Corresponding calculations for arbitrary density matrices one can easily do numerically. Probably, the only common thing one could say regarding separability properties of such arbitrary density matrices is that such properties should be more similar to ones for the b)-class of equally connected density matrices rather to entanglement properties of the a)-class EC-density matrices. Especially, in the most common (i.e. most asymmetric) case of $ D^{N} $-dimensional density matrix  \textit{one may expect very narrow discrete regions of parameters, where entire system is separable depending on different sub-cases(presence or absence of interaction between subsystems, strong or weak mixing in the off-diagonal matrix elements, etc., as it takes place for different sub-cases of the b)-class of equally connected density matrices of the same dimensionality}. At the same time, from such a comparison with b)-class equally connected situation one may notice that separability of such "asymmetric" arbitrary density matrices should be very\textit{ unstable}: small changes in the matrix elements can destroy "separability windows" making the constituent subsystems to be entangled, probably, except the discrete number of well-separated values of certain matrix elements or their products. 

Hence, now one can see that the above detailed study of the applicability of the heuristic causal separability criterion of Eqs.(17-22) to the equally connected "family" of the arbitrary dimensional density matrices - gives predictions being compatible with general physical observations on the quantum ensembles and, as well, with some well-known ppt-based results for low-dimensional density matrices. All this, in turn, proves the validity of the universal causal separability/entanglement criteria of Eqs.(19,22) revealed for the first time in this paper.

\subsection{Further directions of research on the universal causal separability criterion}.

 Obviously, the heuristic general causal method introduced, applied and justified in the above has very reach perspectives of a further development some of which I would like to outline here. Thus, further problems to be addressed can include:

1. Clarification of all interconnections between the universal causal separability criterion of Eqs.(16-23)and mathematical problem which concerns the calculation of eigenvalues for arbitrary matrices of large enough dimensionality. The underlying problem is purely mathematical and probably it can be resolved to some extent numerically. This surely includes also a further more deep study and, probably, also the experimental verification of the applications of heuristic causal criterion (19-22) to different model physical systems considered in this paper (e.g. arrays of qubits and qutrits, systems of coupled quantum oscillators, many-body localized quantum systems, etc.).

2. Further search for different experimentally important composite quantum systems which would be described by means of a)- and b)-classes of equally connected density matrices.

3. Further search for the additional physical arguments (beyond the ones given in this paper), why the entanglement pictures in two a) and b) categories of equally connected density matrices are so different.

4. Applications of obtained causal separability criteria for equally and(or) unequally connected reduced density matrices which appear in quantum detection schemes for real ensembles of multi-state quantum systems including quantum vibrational states \cite{13,15} coupled to many-body quantum systems in the role of quantum detectors for the former \cite{24,25}.

5. Implementations of concrete forms of the universal causal separability criterion obtained to different quantum computation protocols which can be realized on different physical  platforms \cite{22,23,30,34}.      

\section{Conclusions}

  To conclude, the causal aspects of well-known Peres-Horodecki ppt- (positive partial transpose-) separability criterion have been analysed. As the result, it has been established for the first time that partial transpose operation being performed on arbitrary density matrix  corresponds to the local causality reversal in the case where given density matrix encodes separable ensemble of quantum subsystems or, alternatively, to a reversal of a global time arrow in all cases where its constituent quantum subsystems are entangled with each other. This fact leads to two distinct sorts of channel decompositions: for ignorance- and virtual quantum transition probabilities in separable and entangled cases, correspondingly. 

Such the causal approach being developed for the first time allows for the formulation of the most universal form of the separability (or entanglement) criteria of the causal nature for arbitrary density matrices of arbitrary dimensionality in terms of certain fixed explicit combinations of matrix elements of given density matrix. This causal approach to the separability criteria demonstrated for arbitrary density matrices represents a brand new heuristic viewpoint on the underlying temporal quantum evolution of generic density matrices with respect to their entanglement properties. 

The general results obtained in the above on universal causal separability criterion have been then applied to the important wide category of one-parametric equally connected (EC-) density matrices of arbitrary dimensionality in  $ \mathcal{H}_{D}^{\otimes N} $ Hilbert space. Here a number of important new results have been obtained for the separability/entanglement threshold values of the entanglement parameter as functions of ensemble size $ N \geq  2 $ (i.e. on the number of constituent subsystems of the ensemble) and on the dimensionality $ D \geq 2 $ of Hilbert subspace for each constituent quantum subsystem. The latter revealed interesting structure of the entanglement parameter ($ 0 \leqslant p \leqslant 1 $) values with respect to either the separability or entanglement of given EC-density matrix. In particular, it has been established for the first time, that either separability- or entanglement regions in the values of the entanglement parameter $ p $ in EC-quantum ensembles all can "deform", thus, spreading either separability or entanglement on all available entanglement parameter values depending on the dimensionality $ D $ of Hilbert subspace for each quantum subsystem and on the number $ N $ of such subsystems in given quantum ensemble. As well, it was demonstrated that depending on $ D $ and $ N $ in such one-parametric ensembles of EC- quantum subsystems one can observe the appearance of "separability window" in the values of the entanglement parameter $ p $ , while at all values of this parameter beyond such the separability window, the entire quantum system remains entangled. Interestingly, certain dualities between different regimes of inter-parties coupling have been found to take place as well as the unexpected parallels with some known results from the Anderson localisation picture. It was also shown, that general approach developed gives expectable and in many cases intuitively clear predictions on the separability  for many very common types of equally connected quantum ensembles, such as interacting and non-interacting spin chains, arrays of  either decoupled or interacting quantum oscillators, one-dimensional strongly correlated electron systems, etc. 

All the obtained qualitatively new results and predictions all together seem to fill the substantial gap in our understanding of physical reasons behind the separability/entanglement problem for arbitrary density matrices in its most general formulation. On the other hand, a number of completely new results for separability/entanglement thresholds of equally connected density  matrices also can have a widest possible implementation in the majority of quantum computations- and quantum measurements protocols being performed on a variety of platforms which all make use of underlying quantum interacting many-body systems and quantum statistical ensembles of different nature.

This study was supported by the research grants No.09/01-2020 and No.09/01-2021(2) from the National Academy of Sciences of Ukraine.  

\bigskip

\begin{appendix}
\section{One-parametric family of equally connected (EC-) density matrices}

\subsection{Weakly mixing type of a-class  N-free  EC-density matrices }

One can parametrize the a-class of N-free EC-density matrices $ \hat{\rho}^{(a)}_{N}(p) $ in $ D^{\otimes N} $ Hilbert space via following recurrent relation

\begin{eqnarray} 
\nonumber
\hat{\rho}^{(a)}_{n}(p)= \\ \nonumber
\left \lbrace  (1 - \vert p \vert ) \vert  k_{n}  \rangle \langle  k_{n} \vert + \sum_{\bar{k}_{n}=1}^{(D-1)} \frac{\vert p \vert}{(D-1)} \vert  \bar{k}_{n}  \rangle \langle  \bar{k}_{n}  \vert \right \rbrace \\ \nonumber
 \otimes \hat{\rho}^{(a)diag}_{n-1}(p)  \\ \nonumber
+ \sum_{\bar{k}_{n}=1}^{(D-1)} \left \lbrace   \frac{p}{(D-1)}  \vert  \bar{k}_{n}  \rangle \langle  k_{n}  \vert + \frac{p^{\ast}}{(D-1)} \vert  k_{n}  \rangle \langle  \bar{k}_{n} \vert \right \rbrace \\ \nonumber
\otimes \hat{\rho}^{(a)off-diag}_{n-1}(p)\\ \nonumber
\\ 
 k_{n} \neq \bar{k}_{n}; \ \ \ \ \ \ \nonumber \\ 
 \langle  \bar{k}_{n} \vert k_{n}  \rangle = \langle k_{n} \vert \bar{k}_{n} \rangle = 0 \ \ \ \ \ \ \nonumber
\\
 k_{n},=1,..,D; \ \ \ \ \ \ \nonumber  \\ 
  n = 2,..,N; \ \ \ \ \ \   
\end{eqnarray}

where for $ \hat{\rho}^{(a)}_{2}(p) $ of a-class density matrix one has 

\begin{eqnarray} 
\nonumber
\hat{\rho}^{(a)}_{2}(p)=  \\ \nonumber
  (1 - \vert p \vert ) \vert  k_{1}   \rangle  \langle  k_{1} \vert \otimes \vert  k_{2}   \rangle  \langle  k_{2} \vert \\ \nonumber 
 + \sum_{\bar{k}_{1,2}=1}^{(D-1)} \frac{\vert p \vert}{(D-1)^{2}} \vert  \bar{k}_{1}  \rangle  \langle  \bar{k}_{1} \vert \otimes \vert  \bar{k}_{2}  \rangle  \langle  \bar{k}_{2} \vert   \\ \nonumber
+  \sum_{\bar{k}_{1,2}=1}^{(D-1)}\frac{p}{(D-1)^{2}} \vert  \bar{k}_{1} \rangle \langle  k_{1} \vert \otimes \vert  \bar{k}_{2} \rangle \langle  k_{2} \vert \\ \nonumber
 + \sum_{\bar{k}_{1,2}=1}^{(D-1)} \frac{p^{\ast}}{(D-1)^{2}}  \vert  k_{1}  \rangle  \langle  \bar{k}_{1} \vert \otimes \vert  k_{2}  \rangle  \langle  \bar{k}_{2} \vert  \\ \nonumber
\\ 
 k_{1} \neq \bar{k}_{1}; k_{2} \neq \bar{k}_{2} \nonumber \\
 \langle  \bar{k}_{1,2} \vert k_{1,2}  \rangle = \langle k_{1,2} \vert \bar{k}_{1,2} \rangle = 0 \nonumber
\\
 k_{1},\bar{k}_{1}=1,..,D.   \ \ \ \  
\end{eqnarray}

Obviously, in the case $ D=2 $ and $ N=2 $ the a-class N-free EC-density matrix of Eqs.(A1,A2) reduces to density matrix which is equivalent to one from the example by Gisin and Peres \cite{5} . In Eqs.(A1,A2) $ p $ is a parameter (complex, in general) which determines the \textit{degree of entanglement} between any among $ N $ equivalent quantum subsystems and all its environment including another $ N-1 $ counterparts. Without the loss of generality one can parametrize $ 0 \leq \vert p \vert \leq 1 $. Additional pre-factor $ \frac{1}{(D-1)} $ is introduced into definition (A1,A2) of EC-density matrix $ \hat{\rho}^{(D)}_{N}(p) $ in order to provide a proper normalization of density matrix. Notice that chosen normalization corresponds to the model of our system where the amplitudes of quantum transitions between completely orthogonal system's configurations become suppressed in the limit of either $ N\rightarrow\infty $ and $ D > 2 $ or in the limit $ D \rightarrow \infty $, which both describe physically a small fraction of completely orthogonal configurations of subsystems' quantum states in the final measured quantum state of the ensemble. 

The opposite case of substantial fraction of completely orthogonal configurations in the measured quantum state of the ensemble can be realized only due to a very intense "mixing" between different configurations, i.e. due to strong interaction of the entire quantum system with its surrounding on some stages of its temporal evolution. The EC-density matrices of such type differ from ones of Eqs.(A1,A2) by the absence of pre-factor $ 1/(D-1) $ in the off-diagonal elements. We will also consider this particular case in below.   


Now one can see from Eqs.(A1,A2) that the configuration representation (23) of such density matrix is given by Eq.(24) from the main text. Especially, from Eq.(24) it is evident that a-class equally (or Werner-) connected density matrices are invariant with respect to cyclic permutation $ n \rightarrow (n + m) $ with natural  $ 1 \leq m \leq N $ and "periodic boundary condition" of the form $ \vert  k_{N+m}  \rangle = \vert  k_{m}  \rangle  $ and as well with respect to LCR transformation in any $ m $ of its constituent subsystems. Such a sort of "translation invariance" for density matrices (A1,A2) with respect to the index $ n $ which numbers each among $ N $ equal subsystems means by itself the \textit{equal degree of entanglement} of all $ N $ EC-quantum subsystems with each other and with their surroundings. The equal degree of entanglement in density matrices for EC-subsystems is, in turn, a direct consequence of the fact that all $ N $ these subsystems are identical to each other by assumption and hence nothing should change in density matrix elements (16) if one decide to permute (or equally to renumber) any amount of $ m \ (m \leq N) $ identical quantum subsystems in our $ D^{N} $ dimensional quantum system. Therefore, causal entanglement/separability criterion we will find for a-class of EC-density matrices should be valid for any subset of $ m=1,2,..,N-1 $ among $ N $ equivalent  $ D $-dimensional subsystems of the ensemble.  

Without the loss of generality for arbitrary EC-case one can define the entanglement parameter $ p=p(t) $ for given EC-density matrix (where $ t $ is the moment of time all matrix elements of a given EC-density matrix are defined at) as follows

\begin{eqnarray} 
p=p_{j}(t)= \left \langle \mathcal{T}_{{K}} \exp^{\left\lbrace -i \int_{C_{K}(t)}dt' \hat{H}_{prep/meas(j)}(t') \right\rbrace}  \right \rangle_{env}, 
\end{eqnarray}

where the r.h.s. of this equation is a time propagator of the entire quantum system due to preparation and subsequent measurement of a given density matrix or, more precisely, it represents an average over all external degrees of freedom (or equally over environment given quantum system is entangled with) from the Keldysh contour $ C_{K}(t) $ -ordered evolution operator ( or $ T $-exponent) of the entire quantum system. The entire system is supposed to be prepared/evolved till the present moment $ t $ by means of a certain "preparation/measurement Hamiltonian" $ \hat{H}_{prep/meas(j)}(t) $ for any $ j $-th configuration of observable eigenvalues. This Hamiltonian represents certain (strongly decreasing, in general) function of time (in Eq.(A3) $ \hbar $ is set to be equal to one). Hence, according to Eqs.(A1-A3) the case $ \hat{H}_{prep/meas(j)}(t)=0 $ (i.e. $ \vert p \vert=1 $ corresponds to the situation where preparation and measurement of a given density matrix has been performed in such a way that the $ j $-th configuration of subsystem's quantum base states becomes excluded from the corresponding observable measurement outcome. This should signal about the entanglement between all the constituent subsystems of our composite quantum system. Whereas the opposite limit  $ p\rightarrow 0  $ means preparation and subsequent measurement of system's entire quantum state which is asymptotically a product (i.e. completely separable) quantum state of its subsystems. This shows that for EC-density matrices the entanglement degree $ p $  strongly depends on a prehistory of a given composite quantum system. Though as soon as for any real quantum systems the details of such the prehistory are almost unknown, an experimentalist in any EC-case deals only with the concrete values of this parameter being measured which constitute a given EC-density matrix. Therefore, in what follows our goal will be to deduce whether a given EC-density matrix encodes separable or entangled state of given $ N $-partite quantum system prepared conditionally to a specific value of the entanglement  degree parameter $ p $ .

Taking all these observations into account and comparing Eqs.(A1,A2) and Eq.(24) with Eq.(23) one can extract from Eq.(24) all matrix elements needed to calculate the ignorance- and virtual quantum transition probabilities according to definitions of Eqs.(17,18). Thus, by substituting proper matrix elements from $ m $-times pt-transformed Eq.(24) into Eqs.(17,18) and performing the summation in (17,18) over all equal contributions from the completely orthogonal configurations (this summation will result in pre-factors $ (D-1)^{N} $ in Eq.(17) and $ (D-1) $ in Eq.(18), correspondingly) one obtains for ignorance- and virtual quantum transition probabilities for a generic $ D^{N} $-dimensional a-class of $ N $-free weakly  mixing equally connected (EC-) density matrix following expressions

\begin{eqnarray} 
P^{(a.w.)}_{\varnothing}(p)= (1 - \vert p \vert )^{N}\vert p \vert^{N} 
\end{eqnarray}

and

\begin{eqnarray} 
P^{(a.w.)}_{\circlearrowright}(p) = \frac{ \vert p \vert^{2N}}{(D-1)^{2N-1}}. 
\end{eqnarray}

Substituting probabilities (A4,A5) into basic Eq.(19) one arrives Eq.(26) from the main text. Remarkably, the phase of a complex parameter $ p $ and parameter $ m(j) $ both cancel out from the expressions for $ P^{(j)}_{\varnothing}(p) $ and $ P^{(j)}_{\circlearrowright}(p)  $ in the case of a-class of equally connected (or Werner-connected) $ D^{N} $ dimensional density matrices. As well, from Eqs.(A4,A5) it follows that the degree of entanglement for the a-class of $ N $-free weakly mixing EC-density matrices turns out to be the same for any number pt-transformed subsystems since all non-zero off-diagonal matrix elements in Eq.(24) are the same and , hence, the density matrix of Eqs.(A1,A2) is \textit{invariant} with respect to the simultaneous pt-transformation of  $ l $ ($ l=1,..,N-1 $) among its $ N $ subsystems.

\subsection{Strongly mixing type of a-class N-free EC-density matrices}

 "Strongly mixing" limit of a-class EC-density matrices unlike the previous "weakly mixing" case is characterized by configuration-independent equal off-diagonal matrix elements which are also independent on the dimensionality $ D $ of ensemble subsystems. Such the model can describe, for example, any ensemble of  quantum objects which were equally entangled with each other and their environment in the past due to strong uniform interaction between such subsystems via their common surrounding. In other words, this case is characterized by strong "mixing" of all constituent subsystems' quantum states in all averages for observables one could measure and calculate. This type of equal entanglement for the a-class of N-free EC-density matrices of subsystems being decoupled from each other by the moment of density matrix elements measurement -differs only a little from the previous "weakly mixing" case in the mathematical description of the model. However, as we will see in below, surprisingly, the properties of the entanglement in composite quantum systems (or equally, quantum ensembles) of such type become entirely different as compared to a  "weakly mixing" case has been studied in the above. Indeed, in order to obtain formulas for the arbitrary strongly mixing a-class N-free EC-density matrix from the previous weakly mixing type one needs just to make two simple replacements $ \frac{p}{(D-1)} \rightarrow p $ and $ \frac{p^{\ast}}{(D-1)} \rightarrow p^{\ast}$ in all off-diagonal matrix elements from Eqs.(A1,A2) and Eq.(24). Obviously, such the renormalization in Eqs.(A1,A2,A4,A5) to be substituted into Eq.(19) gives formula (37) from the main text.
 
 \subsection{Interacting (N-coupled) a-class of weakly mixing EC-density matrices}

In the above subsections of the Appendix A we considered a situation for equally connected (EC-)density matrices where different subsystems of the ensemble were decoupled from each other at the moment of time for which given density matrix had been defined. Nevertheless, the EC-density matrix of the latter type can still describe entangled ensemble of subsystems since such subsystems could interact with each other and with their environment in the past, e.g. at the stage of given density matrix preparation. Now it is easy to modify all related results of Eqs.(26-28) from the main text for another important (and probably even more common) physical case where all subsystems of the ensemble remain interacting during all the system's (ensemble's) "history".  For simplicity we consider a limiting situation of strong enough interaction between all $ N \geq 2 $ subsystems of given ensemble when eigenvectors $ \vert k_{n} \rangle $ and $ \vert k'_{n'} \rangle $, ($ n \neq n' $) related to the same $ k $-th eigenvalue of certain ensemble's observable $ k $ (we suppose all $ N $ subsystems to be equivalent to each other) in different ($ n $-th and $ n' $-th) subsystems of the ensemble are not orthogonal to each other due to the fact all $ N $ subsystems remain interacting with each other all  the time (for example, due to the non-zero tunnelling between different subsystems). In such the case the true dimensionality of system's Hilbert space reduces sufficiently since the collective strongly correlated quantum states which join all $ N $ subsystems into one $ D $-dimensional subsystem (though being distinct from the initial subsystems of the ensemble) become possible. Thus, we will take (without the loss of generality) for any $ N $-coupled a-class of equally connected density matrices that such a new "effective" dimensionality (or equally the rank of the entire density matrix) is equal to $ D $- i.e. to the number of orthogonal base vectors of states in any among the constituent subsystems of the ensemble. Notice, however, that, generally speaking, this new basis of $ D $ eigenvectors of interacting ensemble is not the same as the corresponding basis for the (N-free) ensemble of the same subsystems being non-interacting with each other. 

Hence, the case of interest corresponds to $ N-coupled $ situation in Eqs.(20,21). Therefore, the modification of the formulas (26-28) for this case remains minimal and concerns only the values of $ N_{o.c.} $ and $ N_{v.t.} $ from Eqs.(20,21) for the N-coupled case under consideration. The result yields (here and below we will use the subscript $ (I) $ to mark all formulas for the fully interacting (N-coupled) case)

\begin{eqnarray} 
P^{(a.w.)}_{\varnothing(I)}(p)= \frac{(1 - \vert p \vert )^{N}\vert p \vert^{N} }{(D-1)^{N-1}}
\end{eqnarray}

and

\begin{eqnarray} 
P^{(a.w.)}_{\circlearrowright(I)}(p) = \frac{\vert p \vert^{2N}}{(D-1)^{N}}. 
\end{eqnarray} 

Now one can obtain Eq.(43) from the main text just substituting Eqs.(A6,A7) into general formula (19).

\subsection{Weakly mixing case of N-free b-class of EC-density matrices}.

Here I consider another important situation of one-parametric equal connection in arbitrary quantum ensembles, where all constituent base states of any configuration of the ensemble can be splitted on two subsets having each one among two distinct probabilities (say, either $ \vert p \vert $ or $ (1-\vert p \vert ) $) to appear in the measurement outcome for certain observable $ k $ in each subsystem of the ensemble. The same holds also for two distinct probabilities of quantum transitions from one ensemble configuration to another within each ensemble's subsystem. Such b-class of EC-density matrices is as common as previous a-class in a number of applications since the latter type describes the ensembles of equal quantum systems with certain quantum states (configurations) of the entire system being discriminated together with respective quantum transitions between them as compared to another configurations of such the ensemble. In what follows it will be convenient to call such another basic type of EC-density matrices as \textit{b-class of EC density matrices}.
In the simplest $ 2 \otimes 2 $ case such the b-class variation of EC-density matrix as well as the related ppt-separability criterion were demonstrated by Horodeckis in their well-known paper \cite{6}. Here we consider different variations of b-class EC-density matrices with respect to weak or strong mixing and interaction for arbitrary $ D^{ \otimes N} $ case. Hence, analogously to Eqs.(A1,A2) the generic case of b-class EC-density matrix for the N-free weakly mixing case reads

\begin{eqnarray} 
\nonumber
\hat{\rho}^{(b)}_{n}(p)= \\ \nonumber
  (1 -  p  )^{m(n)} p^{m'(n)} \vert  k_{n}  \rangle \langle  k_{n} \vert  \otimes \hat{\rho}^{(b)diag}_{n-1}(p) \\ \nonumber 
  + \sum_{\bar{k}_{n}=1}^{(D-1)} \frac{(1 - p )^{m(n)} p^{m'(n)}}{(D-1)} \vert  \bar{k}_{n}  \rangle \langle  \bar{k}_{n} \vert  \otimes \hat{\rho}^{(b)diag}_{n-1}(p)  \\ \nonumber
+ \sum_{\bar{k}_{n}=1}^{(D-1)}   \frac{(1 - p )^{m(n)} p^{m'(n)}}{(D-1)}  \vert  \bar{k}_{n}  \rangle \langle  k_{n}  \vert \otimes \hat{\rho}^{(b)off-diag}_{n-1}(p) \\ \nonumber 
+ \sum_{\bar{k}_{n}=1}^{(D-1)}  \frac{(1 - p )^{m(n)} p^{m'(n)}}{(D-1)} \vert  k_{n}  \rangle \langle  \bar{k}_{n} \vert \otimes \hat{\rho}^{(b)off-diag}_{n-1}(p) \\ \nonumber
\\ 
 k_{n} \neq \bar{k}_{n}; \ \ \ \ \ \ \nonumber \\ 
 \langle  \bar{k}_{n} \vert k_{n}  \rangle = \langle k_{n} \vert \bar{k}_{n} \rangle = 0 \ \ \ \ \ \ \nonumber
\\ \nonumber
 m(n),m'(n)=0,1; \ m(n)\neq m'(n) \\ \nonumber
 k_{n},=1,..,D; \ \ \ \ \ \ \nonumber  \\ 
  n = 2,..,N; \ \ \ \ \ \   
\end{eqnarray}

where for $ \hat{\rho}^{(b)}_{2}(p) $ of b-class density matrix one has 

\begin{eqnarray} 
\nonumber
\hat{\rho}^{(b)}_{2}(p)=  \\ \nonumber
  (1 - p )^{m(n)} p^{m'(n)} \vert  k_{1}   \rangle  \langle  k_{1} \vert \otimes \vert  k_{2}   \rangle  \langle  k_{2} \vert \\ \nonumber 
 + \sum_{\bar{k}_{1,2}=1}^{(D-1)} \frac{(1 - p )^{m(n)} p^{m'(n)}}{(D-1)^{2}} \vert  \bar{k}_{1}  \rangle  \langle  \bar{k}_{1} \vert \otimes \vert  \bar{k}_{2}  \rangle  \langle  \bar{k}_{2} \vert   \\ \nonumber
+  \sum_{\bar{k}_{1,2}=1}^{(D-1)}\frac{(1 - p )^{m(n)} p^{m'(n)}}{(D-1)^{2}} \vert  \bar{k}_{1} \rangle \langle  k_{1} \vert \otimes \vert  \bar{k}_{2} \rangle \langle  k_{2} \vert \\ \nonumber
 + \sum_{\bar{k}_{1,2}=1}^{(D-1)} \frac{(1 - p )^{m(n)} p^{ m'(n)}}{(D-1)^{2}}  \vert  k_{1}  \rangle  \langle  \bar{k}_{1} \vert \otimes \vert  k_{2}  \rangle  \langle  \bar{k}_{2} \vert  \\ \nonumber
\\ \nonumber
 m(n),m'(n)=0,1; \ m(n)\neq m'(n) \\ \nonumber
  k_{1} \neq \bar{k}_{1}; k_{2} \neq \bar{k}_{2} \nonumber \\
 \langle  \bar{k}_{1,2} \vert k_{1,2}  \rangle = \langle k_{1,2} \vert \bar{k}_{1,2} \rangle = 0 \nonumber
\\
 k_{1},\bar{k}_{1}=1,..,D.   \ \ \ \  
\end{eqnarray}

Equations (A8,A9), in turn, imply the configuration representation (23) in the form of Eq.(25) from the main text. Evidently, Eqs.(A8,A9) as well as Eq.(25) allows one to construct the analogs of Eq.(A4,A5) for the ignorance- and virtual quantum transition (to all  configurations being completely orthogonal to the $ j $-th one) probabilities in the following form

\begin{eqnarray} 
P^{(b.w.)}_{\varnothing[m]}(p)= (1 - p )^{2m(j)}p^{2(N-m(j))} 
\end{eqnarray}

and

\begin{eqnarray} 
P^{(b.w.)}_{\circlearrowright[m]}(p) = \frac{(1 - p )^{2(m(j)-m)}p^{2(N-m(j)+m)}}{(D-1)^{2N-1}}, 
\end{eqnarray}

where as well as in Eq.(25) one has $ m=\pm1,\pm2,..,\pm(N-1) $ which corresponds to all changes in powers $ m(j) $ for given $ j $-th configuration which are possible for this configuration due to pt-transformations in the subspaces of a fixed number $ l=1,2,..,N-1 $ of subsystems whose entanglement with each other in the entire ensemble one wants to explore by means of causal entanglement criterion (17-22). Substituting Eqs.(A10,A11) into Eq.(19) yields Eq.(29) from the main text.

\subsection{Interacting (N-coupled) weakly mixing b-class of EC-density matrices}

Analogously to the formulas (A6,A7) for weakly mixing a-class N-coupled EC-density matrices one can obtain separability/entanglement criterion for the b-class of weakly mixing N-coupled density matrices applying Eqs.(20,21) to Eq.(19) with respect to configuration representation (25) for given type of EC-density matrices. Then one can obtain following formulas  for the ignorance-and virtual quantum transition (between completely orthogonal configurations) probabilities

\begin{eqnarray} 
P^{(b.w.)(I)}_{\varnothing[m]}(p)= \frac{(1 - p )^{2m(j)}p^{2(N-m(j))} }{(D-1)^{N-1}}
\end{eqnarray}

and

\begin{eqnarray} 
P^{(b.w.)(I)}_{\circlearrowright[m]}(p) = \frac{(1 - p )^{2(m(j)-m)}p^{2(N-m(j)+m)}}{(D-1)^{N}}. 
\end{eqnarray} 

Substituting Eqs.(A12,A13) into basic Eq.(19) one obtains Eq.(47) from the main text. 

\end{appendix}

\end{normalsize}

\end{document}